\documentclass[epj-spec]{svjour}
\usepackage{graphics}
\usepackage{a4,rotating,latexsym,amsfonts,epsf,psfrag}
\usepackage{amsmath,amssymb}
\usepackage{graphicx}
\usepackage{setspace}

\newcommand{\bea}{\begin{eqnarray}}
\newcommand{\eea}{\end{eqnarray}}
\newcommand{\barr}{\begin{eqnarray}}
\newcommand{\earr}{\end{eqnarray}}
\newcommand{\bas}{\begin{eqnarray*}}
\newcommand{\eas}{\end{eqnarray*}}
\newcommand{\lap}{\mbox{$\bigtriangleup$}}

\newcommand{\ber}{\begin{eqnarray}}
\newcommand{\eer}{\end{eqnarray}}

\DeclareFontFamily{OT1}{rsfs}{}
\DeclareFontShape{OT1}{rsfs}{m}{n}{ <-7> rsfs5 <7-10> rsfs7 <10->rsfs10}{}
\DeclareMathAlphabet{\mycal}{OT1}{rsfs}{m}{n}
\begin{document}
\title{Numerical modeling of black holes as sources of gravitational waves in a nutshell}
\author{Sascha Husa\fnmsep\thanks{\email{sascha.husa@gmail.com}}}
\institute{Institute for Theoretical Physics, University of Jena}
\abstract{
These notes summarize basic concepts underlying numerical relativity and in particular
the numerical modeling of black hole dynamics as a source of gravitational waves. Main topics are
the 3+1 decomposition of general relativity, the concept of a well-posed initial 
value problem, the construction of initial data for general relativity, trapped surfaces 
and gravitational waves. Also, a brief summary is given of recent progress regarding the numerical
evolution of black hole binary systems.
} 
\maketitle

\newcommand{\Scri}{{\mycal J}}
\newcommand{\scri}{{\mycal J}}


\newcommand{\A}{\alpha}
\renewcommand{\S}{\beta}
\renewcommand{\a}{\chi}
\newcommand{\K}{\chi}
\newcommand{\g}{\Gamma}
\newcommand{\spec}[1]{\sigma_{#1}}
\newcommand{\I}{\text{\emph{\textbf{i}}}}
\newcommand{\ecs}[1]{V^{#1}}
\newcommand{\es}[1]{V_{#1}}
\newcommand{\kh}{\hat{k}}

\newcommand{\bigO}{\mathcal O}
\newcommand{\pd}[2]{\frac{\partial #1}{\partial #2}}
\newcommand{\Vr}{\frac{V}{r}}
\def\NP{c_{\scriptscriptstyle\text{NP}}}
\newcommand{\mEXT}{m_{\scriptscriptstyle\text{EXT}}}
\newcommand{\mSSH}{m_{\scriptscriptstyle\text{SSH}}}
\newcommand{\SSH}{{\scriptscriptstyle\text{SSH}}}
\newcommand{\mB}{m_{\scriptscriptstyle\text{B}}}
\newcommand{\mBH}{m_{\scriptscriptstyle\text{BH}}}
\newcommand{\uB}{u_{\scriptscriptstyle\text{B}}}
\newcommand{\tauB}{\tau_{\scriptscriptstyle\text{B}}}

\newcommand{\Lien}{{\cal L}_{n}{}}
\newcommand{\DIII}{\,{}^{\scriptscriptstyle(3)\!\!\!\:}\nabla}
\newcommand{\ROI}{\,{}^{\scriptscriptstyle(0,1)\!\!\!\:}\hat R}
\newcommand{\RII}{\,{}^{\scriptscriptstyle(1,1)\!\!\!\:}\hat R}
\newcommand{\epsIII}{\,{}^{\scriptscriptstyle(3)\!\!\!\:}\epsilon}
\newcommand{\hG}{\mbox{$\hat{G}$}}
\newcommand{\hR}{\mbox{$\hat{R}$}}
\newcommand{\hp}{\mbox{$\hat{\phi}$}}
\newcommand{\hT}{\mbox{$\hat{T}$}}
\newcommand{\Om}{\mbox{$\Omega$}}
\newcommand{\tC}{\mbox{$\tilde{C}$}}
\newcommand{\tG}{\mbox{$\tilde{G}$}}
\newcommand{\tT}{\mbox{$\tilde{T}$}}
\newcommand{\tg}{\mbox{$\tilde{g}$}}
\newcommand{\tn}{\mbox{$\tilde{\nabla}$}}
\newcommand{\tp}{\mbox{$\tilde{\phi}$}}
\newcommand{\tR}{\mbox{$\tilde{R}$}}
\newcommand{\thR}{\mbox{$\hat{\tilde R}$}}
\newcommand{\thT}{\mbox{$\hat{\tilde T}$}}
\newcommand{\thG}{\mbox{$\hat{\tilde G}$}}

\newcommand{\uhat}{\hat u}
\newcommand{\vhat}{\hat v}
\newcommand{\what}{\hat w}
\newcommand{\D}{D^{(1)}}
\newcommand{\DD}{D^{(2)}}
\newcommand{\Dhat}{\hat D^{(1)}}
\newcommand{\DDhat}{\hat D^{(2)}}

\newcommand{\bm}[1]{\mbox{\boldmath$#1$}}               
\def\p{\partial}
\renewcommand{\Re}{{\rm Re}}
\newcommand{\ournote}[1]{\textcolor{red}{\bf #1}}
\newcommand{\ud}{\, \mathrm d}

\newcommand{\columntwo}[2]
{
  \left (
  \begin{array}{c}
    {#1} \\
    {#2}
  \end{array}
  \right )
}

\newcommand{\columnthree}[3]
{
  \left (
  \begin{array}{c}
    {#1} \\
    {#2} \\
    {#3}
  \end{array}
  \right )
}

\newcommand{\matrixtwotwo}[4]
{
  \left (
  \begin{array}{lll}
    {#1} & \quad & {#2} \\
    {#3} & \quad & {#4}
  \end{array}
  \right )
}

\newcommand{\matrixthreethree}[9]
{
  \left (
  \begin{array}{lll}
    {#1}    & {#2}    & {#3} \\
    {#4}    & {#5}    & {#6} \\
    {#7}    & {#8}    & {#9}
  \end{array}
  \right )
}

\section{Introduction}\label{sec:intro}

The theory of general relativity (GR) has enchanted generations 
of physicists through its conceptual and mathematical beauty, 
which is expressed in the simplicity and geometric nature of the Einstein equations,
$$
G_{ab}=8\pi \kappa T_{ab},
$$
relating the curvature of spacetime to its matter content. It is in particular the 
connection of physical concepts such as the equivalence principle to a geometric
description of nature, which is stressed in introductions to the field. 
General relativity  makes many exciting physical predictions, such as
the big bang, black holes, or gravitational waves. The theory also provides a seemingly
 inexhaustible supply of mathematical challenges, and
many deep mathematical insights have been gained in trying to understand the physical content
of the Einstein equations, such as the positive mass theorem \cite{Schoen79b} or the nonlinear
stability of Minkowski space \cite{Christodoulou93}. The currently emerging fields of gravitational
wave astronomy, or, more generally, general relativistic astrophysics, are however changing this
picture on a rapid timescale.
A large international effort is underway to establish
a network of gravitational wave detectors, such as LIGO
\cite{Abramovici92,LIGO_web},
GEO \cite{GEO1,GEO_web} and VIRGO \cite{VIRGO_web}, some of which are already taking data at
design sensitivity, and first publications have set new upper limits on the radiation
from several sources (see {\it e.g.} 
\cite{Abbott:2007xi,Abbott:2007wu,Abbott:2007ce}).

The field where the contact between mainstream general relativity and new applications
is perhaps most intense, is numerical relativity (NR). Here one tries to 
systematically explore the solution space of the Einstein equations with numerical methods.
One of the most advertised goals is to model the inspiral and collision of two black holes,
and the resulting gravitational wave signals, in order to provide templates for
gravitational wave data analysis. Many other applications are no less
exciting, {\it e.g.}
the study of cosmological singularities ({\it i.e.} the predictions of general relativity
about the very early universe), or critical collapse, where one probes the 
regime of extremely high curvatures created from the collapse of matter
fields, {\it  i.e.}
a regime where classical general relativity will ultimately lose its validity and quantum
effects will play a role.

While one of the most timely challenges for numerical relativity is the establishment of
a ``data analysis pipeline'', connecting analytical calculations
of the early inspiral phase with numerical simulations and gravitational wave searches
in actual detector data, on a technical level such simulations lead to challenging questions
not only regarding the physics of general relativity, but also in applied mathematics, 
high performance computing and software engineering. 

In order to set up reliable and efficient algorithms that predict detector signals
from ``first principles''-type numerical solutions of the Einstein equations it is essential
that the individual steps are clearly understood, and subtleties or open issues be made
transparent. Already the concept of gravitational wave emission from some source is
highly nontrivial in general relativity, and touches upon the global structure of spacetimes.
Setting up the Einstein equations as an initial value problem and solving them numerically
raises several new problems in the theory of partial differential equations and their
numerical solution.
The complexity of the Einstein equations requires a sophisticated approach toward
writing software in order to produce reliable results. Efficiency is paramount, when
one actually aims to make the connection to data analysis, where large parameter studies
are expected to be required. Understanding optimal choices in this regard requires 
a careful analysis of the interplay of the continuum equations, numerical methods and
their implementation on current generations of high performance computing equipment.

A key idea when one wants to study the solutions of the Einstein equations in a systematic way,
is to write the Einstein equations in the form of a well posed initial value problem. That this
is possible, is not trivial, and an important test of the theory. Like the other fundamental theories
of interactions (the electroweak interaction and quantum chromo-dynamics), general relativity can be interpreted
as a gauge theory (in this case of the diffeomorphism group), and initial value formulations thus
contain constraint equations.
These constraint equations, which are associated with the diffeomorphism invariance, and the general
subtleties of diffeomorphism invariance, like the absence of a fixed spacetime background which
is essential in most areas of physics, create challenging problems for the mathematical treatment
of the classical theory, its quantization, and for any treatment with approximation methods, such
as numerical simulations. In a sense, the root of the problems to quantize gravity, is the same
as for many problems to solve the Einstein equations numerically.

Discretizing GR can be approached in very different ways:
\begin{itemize}
\item One approach, that also has found applications in quantum gravity, is the direct 
discretization of the geometry ({\it e.g.} 
in Regge calculus \cite{Gentle-2002:Regge-calculus-review}).

\item A related idea is to use ``discrete differential forms'', in analogy of a very successful
method in electromagnetism, see {\it e.g.} \cite{Meier:2003bn}.

\item The mainstream approach, which I will follow in these notes, is to view the Einstein equations
as a standard PDE (partial differential equations) problem.
\end{itemize}
All of the approaches however share the fundamental problem that 
due to the presence of  the rather complicated constraints
we have no direct access to physical degrees of freedom.
As a consequence approximate solutions of constrained systems may not manifestly
satisfy the constraints, it turns out that exponential drift off the constraint surface is
typical, as is illustrated by the Maxwell example of eq. (\ref{eq:maxwell_contraint_prop}).
It is possible to trade evolution equations for constraints when constructing
the solution -- a minimal number of (hyperbolic) evolution equations
would correspond to the local degrees of freedom ({\it i.e.} two for vacuum gravity).
This may improve the quality of numerical evolutions, but there is no guarantee: since there
are more equations than evolution variables, one still needs to worry whether the evolution
equations that are not used to construct the solution will be (manifestly) satisfied.

The good news for numerical relativity is of course that
in contrast to perturbation techniques numerical simulations
do not rely on the smallness of physically relevant parameters, and can thus be applied
in very general situations.
Carefully crafted numerical simulations provide error estimates and convergence results.
A particular challenge for the field, for instance, is to clarify the range of
validity of post-Newtonian (PN) expansions \cite{Blanchet02}, and significant progress has been
made in this direction \cite{Buonanno06imr,Baker:2006ha,Berti:2007fi,Pan:2007nw,Hannam:2007ik}.

\section{Gravitational Waves}


Like electromagnetism, general relativity predicts waves:
aspherically accelerated masses emit gravitational
radiation. Both Maxwell's and Einstein's theories have two local degrees of freedom, which
correspond to the two polarization states of waves in the theory.
While in electromagnetism waves carry no monopole moment, in relativity there are
no monopoles {\em and} no dipoles, to leading order wave emission is thus quadrupolar.

Under ``everyday conditions'', including the conditions in a gravitational wave
detector, the amplitude of gravitational waves is extremely small, and it is sufficient
to consider the linearized theory.
Consider the approximation $g_{ab}=\eta_{ab}+h_{ab}$ for
weak gravitational fields, where $\eta_{ab}$ is the metric of Minkowski spacetime,
and $h_{ab}$ is a small perturbation.
Using the definition
$$
\bar h_{ab}=h_{ab}-\frac{1}{2}\eta_{ab} {h^{c}}_{c}
$$
one can show that
$$\Box \bar h_{ab}=0$$
if the gauge condition
$$
\partial \bar h_{ab}/\partial x^b=0
$$
is satisfied. The tensor $\bar h_{ab}$ thus satisfies the standard wave equation
on a flat background. 
It can be shown that a further gauge condition can be adopted to also
set the trace of $h_{ab}$ to zero, $\eta^{ab}\bar h_{ab}=0$.
In this `` transverse-traceless (TT) gauge'', the metric perturbation $h_{ab}$
corresponding to a wave traveling in the $z$-direction of a Cartesian coordinate system
can be written as a linear combination of two polarizations
$h_+$ and $ h_\times$:
\begin{equation}
  \label{eq:hpluscross}
  h_{ij}  = h_+ (\mathbf{e}_{+})_{ij} +
  h_\times(\mathbf{e}_\times)_{ij},
\end{equation}
where $\mathbf{e}_{+,\times}$ are the basis tensors 
\begin{equation}
  \label{eq:basistensors}
  (\mathbf{e}_+)_{ij} = \hat x_i \hat x_j -
  \hat y_i \hat y_j\,, \qquad \textrm{and} \qquad
  (\mathbf{e}_\times)_{ij} = \hat x_{i}\hat y_{j} + \hat x_{j}\hat y_{i}\,,
\end{equation}
and $\hat x$ and $\hat y$ are the unit vectors in the $x$ and
$y$ directions respectively. For an extensive introduction to the linearized
theory of gravitational wave, including the principles of their detection, see \cite{Flanagan:2005yc}.

When modeling sources of gravitational waves in full general relativity, it turns out that
a notion of gravitational waves becomes highly nontrivial: because of diffeomorphism invariance
there is no background spacetime available, on which gravitational waves can be defined
in an unambiguous way. Such a definition is only possible asymptotically for isolated systems.
When the geometry ``flattens out'' to approach Minkowski spacetime at large distance from
the sources, then  Minkowski spacetime can be used as a suitable background,   see
sec. \ref{sec:energy}.

The weak field approximation can also be used to compute the
radiation from sources, {\it e.g.} accelerated bodies.
For velocities $v \ll c$ and for wavelengths $\lambda$ much larger than the size of the system,
a weak field calculation yields the loss of energy of a system of massive bodies as
\begin{equation}\label{eq:quadrupole}
\frac{dE}{dt}=-\frac{G}{5c^5}\sum_{i,j}\left(\frac{d^3Q_{ij}}{dt^3}\right)^2
\end{equation}
with $Q_{ij}=\int\varrho(x_ix_j-\frac{1}{3}\delta_{ij}r^2)d^3x$ is the
mass quadrupole moment. Eq. (\ref{eq:quadrupole}) is known as the quadrupole formula
(for the energy loss of the system).
The radiation power scales with the sixth power of
the frequency of the system.
Due to the weakness of gravity, expressed in the factor $\frac{G}{5c^5}$, thus
only systems of astrophysical dimensions
-- large masses moving at a significant fraction of the speed of light
-- generate significant amounts of gravitational radiation.
While the motion of the earth around the sun only generates around 200 Watt of gravitational
radiation, a close neutron star binary (100 km separation, 100 Hz)
would generate approximately $10^{45}$ Watt.
In the inspiral and collision of two black holes, approximately 4 \% of the total energy is radiated
away in the form of gravitational waves -- for the collision of two stellar size black holes with ten solar
masses each, this
would already be around $0.8$ solar masses!
Indirect confirmation of the predictions of general relativity for the
energy loss of a binary system due to the emission of gravitational wave has been possible
by measuring the tightening of the orbit of the Hulse-Taylor binary pulsar \cite{HulseTaylor74,Taylor82a}.

Two essential properties of gravitational waves for astrophysical observations
are that they are not shielded by interstellar masses,
and that detectors are sensitive to amplitude rather than intensity.
The first means that not only will gravitational wave observations render possible
the direct observation of phenomena that have not been observable so far,
but we will also be able to see regions that have been hidden behind dust clouds.
Gravitational waves represent the bulk motion of fast compact objects instead of
incoherent superposition of many atoms, which is typical of electromagnetic observations.
The observation of coherent wave trains makes gravitational wave observations similar to
detecting sound waves, rather than to producing images, and in particular
results in a direct sensitivity to amplitude. Consequently the
sensitivity scales with $1/distance$ instead of $1/distance^2$ as for electromagnetic
observations.
This makes it much easier to observe objects at extreme distances, and also means
that any improvement in detector sensitivity by some factor $\lambda$ yields a factor
$\lambda^3$ in event rates.

 \section{The Initial Value Problem for General Relativity}

\subsection{The Maxwell Equations}


GR is a classical relativistic field theory much like Maxwell's
theory of electromagnetism, but it is nonlinear, and its dynamics involves
the very structure of spacetime itself.
We first discuss the structure of the Maxwell equations as an example.
The Maxwell equations can be elegantly and geometrically formulated in a 4-dimensional
way as
\begin{equation}\label{eq:Maxwell}
dF = 0, \qquad d*F = 4 \pi *J,
\end{equation}
where $F$ is the Faraday tensor of electromagnetic field strengths, and $J$ is the 4-dimensional
current. For practical purposes it is useful to solve the Maxwell equations as an initial value
problem, that is to specify suitable data at some instant of time, 
and then evolve them to the future. The formulation as an initial value problem allows a
systematic approach to constructing solutions: First one needs to classify permissible
initial data, then the future and/or past development of such data is determined uniquely by
evolution equations. The Maxwell equations (\ref{eq:Maxwell}) are not written
in this form, in particular it is not immediately obvious what permissible initial data would be,
and whether time evolution of such data prescribes a unique solution.

A standard way to proceed is to prescribe a foliation of spacetime, {\it e.g.} by specifying a time
function $t(x^\mu)$, such that the surfaces $\Sigma_t$, defined by $t=const.$, form
a 1-parameter foliation of spacelike surfaces. One can then perform a geometric split
of  4-dimensional tensorial quantities into 3-dimensional objects ({\it i.e.} objects defined
in the tangent and co-tangent space of $\Sigma_t$), 
and consider the equations for these 3-dimensional objects.
It will prove useful to define a timelike unit normal to the surfaces $\Sigma_t$,
$$
n_a := \frac{\nabla_a t}{\nabla^c t \nabla_c t}.
$$
The timelike unit normal can be used to define projection operators onto timelike (vertical)
and spacelike (horizontal) directions. It is easy to check that $n^a n_b$ projects onto
the direction of the timelike unit normal, and  the induced metric 
\begin{equation}\label{eq:first_fund_form}
h_{ab} = g_{ab} + n_a n_b
\end{equation}
projects
onto the tangent space of the spacelike hypersurface $\Sigma_t$, where $g_{ab}$ denotes the
(Lorentzian) spacetime metric, and $h_{ab}$ then is the induced Riemannian (positive
definite) metric on $\Sigma_t$.
A covariant derivative $D_a$ on $\Sigma_t$ that is compatible with the metric
$h_{ab}$, $D_a h_{bc} = 0$, can be defined as 
$$
D_{c}T^{a_{1}\dots a_{r}}_{b_{1}\dots b_{s}}:=
h^{c'}_{c}h^{a_{1}}_{a'_{1}}\dots h^{a_{r}}_{a'_{r}}
h^{b'_{1}}_{b_{1}}\dots h^{b'_{s}}_{b_{s}}
\nabla_{c'}T^{a'_{1}\dots a'_{r}}_{b'_{1}\dots b'_{s}}.
$$
Excellent pedagogical discussions of this geometric splitting procedure can be found
in \cite{Hawking73a,Wald84}.

For the Maxwell equations one can now define
$$
E^a = F_{ab} n^b, \qquad B^e = \frac{1}{2} F_{cd} {h^{c}}_{a} {h^{d}}_{b} \epsilon^{abef} n_f,
$$
the electric and magnetic field as projections of the field strength tensor $F_{an}$, 
as well as the charge density $\rho = J^a n_a$ and current
$j_b = J^a h_{ab}$.
On Minkowski spacetime, and also using the standard slices of Minkowski spacetime,
the Maxwell equations separate into the
following two groups of equations in terms of the electric and magnetic fields:
\begin{equation}
D_a E^a = 4 \pi \rho, \qquad
D_a B^a = 0 \label{eq:EB_cons},
\end{equation}
and
\begin{eqnarray}
\partial_t E^a = \epsilon_{abc}\partial^b B^c - 4 \pi j_a, \label{eq:E_evol}\\
\partial_t B^a = -\epsilon_{abc}\partial^b E^c \label{eq:B_evol}.
\end{eqnarray}
Since Eqs. (\ref{eq:EB_cons}) do not contain time derivatives, they
have to be interpreted as constraint equations that restrict the space of allowed
initial data, while (\ref{eq:E_evol},\ref{eq:B_evol}) are evolution equations.
It is not hard to show that if the constraints are satisfied initially, they will be satisfied
for all times, furthermore, the evolution equations determine the time evolution
in a unique way. Without these two facts, the initial value description of the Maxwell equations,
eqs. (\ref{eq:EB_cons}-\ref{eq:B_evol}) would make little sense.
Because of their linearity, the discussion of the initial value problem, and the solution of the 
constraints are tremendously
simpler for the Maxwell equations than for the Einstein equations, but the essential structure
is the same.
Some remarks are in order: Counting degrees of freedom in
eqs. (\ref{eq:EB_cons}-\ref{eq:B_evol}),
we find that there are 2 local degrees of freedom (6 first order evolution equations minus
2 constraint equations makes 4 first order in time or 2 second order in time
equations), these correspond to the two polarization states of the electromagnetic
field. It is clear that while for the continuum solution the constraints propagate (they
are satisfied at all times by virtue of the evolution equations if they are satisfied at
the initial time), this is not necessarily true for a numerical implementation. Discretization
error may give rise to a growth in the constraints, which may lead to instability or are least
spoil the physical validity of the approximate solution. We will see that the same problem
appears for the Einstein equations. In electromagnetism, many approaches have been
developed to deal with this issue, see \cite{Toth2000} for a comparison. The most
standard approach uses a formulation of the Maxwell equations in
differential forms language, and directly translates the differential forms concepts to
the discrete level \cite{Bossavit1998}. This way, the constraints can be solved manifestly in the discrete
problem.

The problems of preserving the constraints of the Maxwell theory become more pronounced in curved
spacetime, and already in curved slices of Minkowski spacetime, where
the constraint propagation equations become
\begin{equation}\label{eq:maxwell_contraint_prop}
{\cal L}_{n} D_i E^i  =  - K D_i E^i , \qquad {\cal L}_{n} D_i B^i =   - K D_i E^i
\end{equation}
where ${\cal L}_{n}$ is the Lie derivative along the timelike unit normal of the hypersurface,
and $K = D_a n^a$ is the trace of the extrinsic curvature of the hypersurface (see below). 
Here the sign of $K$ is chosen such that $K>0$ corresponds to expansion, and $K<0$ to collapse. Clearly,
in the collapsing case, initial violations of the constraints will be amplified.
While a simple redefinition of the fields (densitizing them) 
solves the problem in the Maxwell case \cite{Anton05}, this still provides
a valuable toy model for similar problems in numerical relativity \cite{Husa:2005ns}.

\subsection{The Einstein Equations}

The Einstein equations $G_{ab}=8\pi \kappa T_{ab}$
can be written as an initial value problem in formal analogy with the previous
discussion of the Maxwell equations. There is much freedom in how to do this,
and the route we will take is to very briefly introduce the most traditional way,
and then to discuss its issues with regard to numerical relativity.

First, we need to make a careful choice of the topology of our spacetime manifold $M$,
since the Einstein equations allow solutions that violate causality in a global way,
{\it i.e.} geometries where there exists no spatial hypersurface $\Sigma_t$ that uniquely determines
the geometry at all earlier and later times. A necessary requirement for causality \cite{Wald84}
is that the topology of the spacetime be $\Sigma \times I$, where
$\Sigma$ is a three-dimensional manifold, and $I$ is a (possibly infinite) interval. 
We can then choose coordinates
such that the line element takes the form
\begin{equation}\label{eq:ds_3+1}
ds^2 = -\alpha^{2} dt^{2} + \gamma_{ab} (dx^{a} + \beta^a dt) (dx^{b} + \beta^b dt).
\end{equation}
In particular we have thereby singled out a time function $t$.

We can now write the Einstein equations in terms of 3-dimensional objects:
the induced metric $h_{ab}$ (the definition in (\ref{eq:ds_3+1}) coincides with the definition
(\ref{eq:first_fund_form}), the shift vector $\beta^a$ and the lapse function $\alpha$.
The coordinate basis vector $(\partial/\partial t)^a$ can be written in terms of the  unit normal $n^a$,
and the lapse and shift as
$$
\left(\frac{\partial}{\partial t}\right)^a = \alpha n^a + \beta^a, \qquad  n^a \beta^a = 0.
$$
A reduction to first order in time of the field equations can be obtained
by introducing the extrinsic curvature (or second fundamental form)
\begin{equation}\label{def_K}
K_{ab}:= \frac{1}{2}{\cal L}_{n}\gamma_{ab} =\frac{1}{2 \alpha}
\left(\dot \gamma_{ab}-D_{a}\beta_{b}-D_{b}\beta_{a}\right),
\end{equation}
${\cal L}_{n}$ is again the Lie derivative with respect to the vector field $n^a$.
Making a particular choice of adding constraints to the evolution equations
one arrives at a ``standard form'' of the 3+1--decomposed Einstein equations,
which has dominated numerical relativity for several decades, and is 
commonly referred to as the ADM-equations, and called the York-ADM equations here, 
since these equations
can be viewed as a variant \cite{York79} of an evolution system discussed by 
Arnowitt, Deser and Misner in \cite{Arnowitt62}:
\begin{equation}\label{eq:ADMconstraints}
D_{b}{K^{b}}_{a} - D_{a}{K^{b}}_{b} = 0 ,\qquad {^{3}R}+({K^{a}}_{a})^{2}-K_{ab}K^{ab} = 0 ,
\end{equation}
\begin{equation}\label{eq:ADMevol}
\dot K_{ab}= -D_{a}D_{b}\alpha+\beta^{c}D_{c}K_{ab}+K_{cb}D_{a}\beta^{c}
-K_{bc}D_{a}\beta^{c} + \alpha({^{3}R_{ab}}+{K_{c}}^{c}K_{ab}),
\end{equation}
where for simplicity matter terms have been set to zero. The first two equations
can be interpreted as constraints analogous to the Maxwell divergence constraints,
while the third equation together with the definition of the extrinsic curvature (\ref{def_K})
forms a first order in time evolution system.
The Bianchi identity $\nabla^a G_{ab}=0$ implies that the constraints propagate, that is they
are satisfied by virtue of the evolution equations at all times if they are satisfied initially.
For a much more detailed discussion of the 3+1 decomposition of the Einstein equations see
\cite{Gourgoulhon:2007ue}.

The big advantage of the ADM equations is their {\em relative} simplicity.
 The problem for numerical 
relativity, which has only been realized after several decades, is that the ``free evolution problem'',
where the constraints are only solved initially and only the evolution equations are used
to construct the solution at later time, is only  weakly hyperbolic in the language of
hyperbolic partial differential equations, and the initial value problem therefore ill-posed
(with the exception of certain subclasses of initial data such as spherical symmetry, which
has added further to the confusion). The issue is discussed in detail in \cite{Calabrese:2005ft}.

\subsection{BSSN: the workhorse formulation for the binary black
hole problem}

For a long time, the standard choice of variables for writing the Einstein
equations was based on the  York-ADM equations
\cite{York-1979-in-Yellow}.
It is known now, however, that the hyperbolic subsystem thus obtained
is not well posed; specifically, it leads to a weakly hyperbolic
set of equations (see {\it e.g.}~\cite{Calabrese:2005ft}). 
A formidable industry of creating improved evolution systems
has produced numerous alternatives, the most popular being 
the ``BSSN family'' \cite{Baumgarte99,Shibata95,Alcubierre2000}.
This formulation is characterized by introducing a contracted connection
term as a new variable, a conformal decomposition of the
metric and extrinsic curvature variables, and adding constraints
to the evolution equations.

Detailed discussions of well-posedness for the BSSN family have
been given by Gundlach and Martin-Garcia \cite{Gundlach:2004jp,Gundlach:2005ta,Gundlach2006}.
The set of evolved variables are the conformally rescaled unimodular three-metric $\tilde \gamma_{ij}$,
the logarithm of the conformal factor $\phi$,
the trace of the extrinsic curvature $K$, the conformally rescaled
traceless extrinsic curvature $\tilde A_{ij}$, and the contracted
Christoffel symbols $\tilde \Gamma^i$:
\begin{eqnarray}\label{BSSNvars}
\phi                & = & \frac{1}{12}\log ({\rm det}\gamma_{ij}), \\
\tilde{\gamma}_{ij} & = & e^{-4\phi}\gamma_{ij},  \\
                  K & = & \gamma^{ij}K_{ij}, \\
\tilde{A}_{ij}      & = & e^{-4\phi}(K_{ij} - \frac{1}{3}\gamma_{ij}K), \\
\tilde{\Gamma}^i    & = & \tilde{\Gamma}^i_{jk}\tilde{\gamma}^{jk}.
\end{eqnarray}
As is usual, we will adopt the convention that indices of
densitized quantities (denoted with a tilde) are raised
and lowered with the conformally rescaled three-metric $\tilde \gamma_{ij}$.
The introduction of new variables leads to new constraints, one differential and two algebraic:
\begin{equation}
{\cal G}=\tilde{\Gamma}^i-\tilde{\gamma}^{jk}\tilde{\Gamma}^i_{jk} = 0,\qquad
   {\cal S} = \det \gamma_{ij} - 1 = 0, \qquad
   {\cal A} = \tilde A^i_i     = 0,
\end{equation}
which are again propagated by the evolution equations. 

The standard Hamiltonian and momentum constraints of general relativity take the form
\cite{Shinkai02a}
\begin{eqnarray*}
{\cal H} &=& e^{-4\phi}\left(\tilde{R} -8\tilde{D}^j\tilde{D}_j\phi
-8(\tilde{D}^j\phi)(\tilde{D}_j\phi)\right)
+\frac{2}{3}K^2
-\tilde{A}_{ij}\tilde{A}^{ij}
-\frac{2}{3} {\cal A} K,
\label{BSSN_ham} \\
{\cal M}_i &=&
 6\tilde{A}^j{}_{i}(\tilde{D}_j \phi)
-2{\cal A}(\tilde{D}_i \phi)
-\frac{2}{3} (\tilde{D}_i K)
+\tilde{\gamma}^{kj}(\tilde{D}_j\tilde{A}_{ki}).
\label{BSSN_mom}
\end{eqnarray*}
The BSSN evolution equations, which are obtained from the Einstein equations
by using the definitions (\ref{BSSNvars})
and making a standard choice for adding constraints, become
\begin{eqnarray*}
{\cal L}_{n}\phi &=& -\frac{\alpha K}{6}, \label{BSSNeqmPHI} \\
{\cal L}_{n}\tilde{\gamma}_{ij} &=& -2\alpha\tilde{A}_{ij},\label{BSSNeqmtgamma}\\
{\cal L}_{n} K &=& -D^iD_i\alpha
+\alpha \tilde{A}_{ij}\tilde{A}^{ij} + \frac{\alpha K^2}{3}, \label{BSSNeqmK} \\
{\cal L}_{n} \tilde{A}_{ij} &=&
-e^{-4\phi}(D_iD_j\alpha)^{TF}
+ \alpha \left(e^{-4\phi} (R_{ij})^{TF}
+K\tilde{A}_{ij}
-2 \tilde{A}_{ik}\tilde{A}^k{}_j\right), \label{BSSNeqmTA} \\
{\cal L}_{n} \tilde{\Gamma}^i &=& -2(\partial_j\alpha)\tilde{A}^{ij} +2\alpha
\big(\tilde{\Gamma}^i_{jk}\tilde{A}^{kj} -\frac{2}{3}\tilde{\gamma}^{ij}(\partial_j K)
+6\tilde{A}^{ij}(\partial_j\phi) \big) \label{BSSNeqmTG},
\end{eqnarray*}
where $\tilde{D}_i$ is covariant derivative associated
with $\tilde{\gamma}_{ij}$,
 ${\cal L}_n = \partial_t - {\cal L}_{\beta}$ is the Lie derivative along the
unit normal, $T_{ij}^{TF}$ denotes the trace-free part of a tensor $T_{ij}$.
The Ricci curvature in terms of the BSSN variables takes the form
\begin{eqnarray*}
R_{ij} &=&  -2\tilde{D}_i\tilde{D}_j\phi
-2\tilde{\gamma}_{ij}\tilde{D}^k\tilde{D}_k\phi
+4(\tilde{D}_i\phi)(\tilde{D}_j\phi)
-4\tilde{\gamma}_{ij}(\tilde{D}^k\phi)(\tilde{D}_k\phi),
\\
 & &
-(1/2)\tilde{\gamma}^{lk}\partial_{l}\partial_{k}\tilde{\gamma}_{ij}
+\tilde{\gamma}_{k(i}\partial_{j)}\tilde{\Gamma}^k
+\tilde{\Gamma}^k\tilde{\Gamma}_{(ij)k}
+2\tilde{\gamma}^{lm}\tilde{\Gamma}^k_{l(i}\tilde{\Gamma}_{j)km}
+\tilde{\gamma}^{lm}\tilde{\Gamma}^k_{im}\tilde{\Gamma}_{klj}.
\end{eqnarray*}
The algebraic constraints are typically solved at every time step, {\it e.g.}
by setting
$\tilde{\gamma}_{ij} \to
{{\tilde\gamma}_{ij}} {\det {\tilde\gamma}^{-1/3}}$ and
$\tilde A_{ij} \to \tilde A_{ij}
- \frac 1 3 \tilde A_{lm} \tilde \gamma ^{il}
\tilde \gamma^{jm}$.

Currently, the standard choice for evolving the lapse function for BSSN-evolutions
is given by the so-called Bona-Masso family of slicing conditions:
\begin{equation}
\partial_t \alpha = - \alpha K f(\alpha),
\end{equation}
in particular the choices $f=1$, corresponding to
harmonic time slicing, and $f=2/\alpha$, which is usually termed
``1+log'' slicing, see \cite{Alcubierre02a}.

In order to obtain long-term stable numerical simulations, it is equally important
to construct a suitable shift vector field $\beta^i$. Here we report on
evolutions where we evolve the shift vector according to a Gamma-freezing prescription
\cite{Alcubierre02a}. A key feature of this particular choice is to drive
the dynamics of the variable $\tilde{\Gamma}^i$ towards a stationary state. As a
``side effect'' this choice creates a coordinate motion that drags the black
holes along an inspiral orbit. A crucial effect of this method is that the resulting coordinate
motion which corresponds to the naive physical intuition reduces artificial distortions in the
geometry, which otherwise could easily trigger instabilities

In summary, the solution procedure for the equations is as
follows.
First, we specify
free data motivated by quasi-equilibrium arguments, then solve
the 9 components of the constraint equations
$({\cal H},{\cal M}_i,{\cal G}^i,{\cal A},{\cal S})$ to obtain
initial data for the 17 evolution variables
($\phi,\tilde{\gamma}_{ij}$,
$K$,$\tilde{A}_{ij}$,$\tilde{\Gamma}^i$).
The evolution system is completed by specifying evolution equations for the
four gauge quantities ($\alpha, \beta^i$), which yields a hyperbolic system that
is second order in space and first order in time, and which determines
the evolution of all 21 components of the ``state vector'' describing
the geometry of spacetime.

In finite difference codes for the solution of nonlinear hyperbolic equations,
it is common practice to add artificial
dissipation terms to all right-hand-sides of
the time evolution equations, schematically written as
\begin{equation}
\partial_t {\bf u} \rightarrow \partial_t {\bf u} + Q {\bf u}.
\end{equation}
Such terms damp out high-frequency noise, {\it e.g.} as produced by mesh-refinement boundaries, and 
can be necessary to guarantee numerical stability for nonlinear problems  \cite{Kreiss73}.
The typical form of this terms is the 
Kreiss--Oliger dissipation operator \cite{Kreiss73} ($Q$) of order $2r$
\begin{equation}
Q = \sigma (-h)^{2 r - 1} (D_+)^{r} \rho (D_-)^{r}/2^{2r},
\end{equation}
for a $2r -2$ accurate scheme, with $\sigma$ a parameter
regulating the strength of the dissipation, and $\rho$ a weight function
that we typically set to unity.

Discretization in space is performed with standard second-, fourth- 
\cite{Zlochower2005:fourth-order,Bruegmann:2006at} or sixth-order \cite{Husa:2007ec}
accurate stencils.
In particular, symmetric stencils are used, with the exception of the
advection terms associated with the shift vector, asymmetric upwind stencils are used.
Time integration is performed by standard Runge-Kutta type methods,
in particular 3rd and 4th order Runge-Kutta and second order accurate
three-step iterative Crank-Nicholson integrators as described in
\cite{Calabrese:2005ft}, where
Courant limits and stability properties are discussed for the types of
equations used here.

\section{Partial differential equations:
Well-posedness and numerical stability for initial value problems}

An excellent and seemingly trivial starting point for a discussion of numerical approximations is
the model problem $F(x,y) = 0$. An important issue in the context of approximations is
the sensitivity in the dependence of a solution $y$ on an input parameter $x$.
It is useful to define a {\em condition number}, which quantifies the worst possible effect on 
$y$ when $x$ is perturbed.
Consider the perturbed eq.  $F(x + \delta x ,y + \delta y) = 0$, and define
$$
K = \sup_{\delta x}
 \frac{\vert\vert \delta y\vert\vert/\vert\vert y \vert\vert}
      {\vert\vert \delta x\vert\vert/\vert\vert x \vert\vert}\ .
$$
If $K$ is small we call the problem well conditioned, if $K$ is large ill conditioned,
and if $K(y)=\infty$ ill-posed or unstable. As a starting point for numerical relativity we clearly 
need the continuum problem to be well-conditioned -- in GR this is far from
trivial, {\it e.g.} because
this will rely on a judicious choice of coordinate gauge.

We call an evolution problem well posed if a unique solution exists 
(in a gauge theory such as general relativity this requires
a gauge choice) and depends continuously on the initial data. The latter condition is usually 
expressed as the condition
\begin{equation}\label{wellposed}
\vert\vert u(t) \vert\vert \leq K e^{a \, t} \vert \vert u(0) \vert\vert,
\end{equation}
where the exponential term ensures robustness with respect to lower order terms, and the constants
$a$ and $K$ can be chosen independently of the initial data.
Note that this condition is only required  local in time, since global in time solutions may not exist in a 
nonlinear theory -- singularities may form!
An obvious crucial question is which norm $\vert\vert . \vert\vert $ is appropriate
for defining well-posedness of a certain type of differential equation. It turns out that
for first order in space and time systems the standard $L_2$-norm is sufficient.
Since the Einstein equations take the form of second differential order equations in the metric,
a complete reduction to first order may seem artifical.
Note that there is an important difference between first order reductions in space and in time:
Reduction to first order in time leads to new evolution equations, but
reduction to first order in space leads to new evolution {\em and}
constraint equations. This enlargement of the phase space may not only reduce
computational efficiency, but can also give rise to instabilities or pathologies
not inherent in the original problem.

First order in time formulations provide a ``normal form'' for ordinary differential equations
and are thus also  convenient for systems of PDEs, which 
in numerical analysis are often discussed in terms of the ``method of lines''. In this
approach first only space
is discretized, and time left continuous. PDEs are thus
converted to coupled systems of ordinary differential equations.
From a physical point of view, first order in time formulations are
attractive, {\it e.g.} because
they are most easily integrated into a Hamiltonian formulation.

The concept of well-posedness translates straightforwardly to the concept of
numerical stability for discrete iterative problems. Consider a simple stable iterative
algorithm
$$
v^{n+1} = Q(t_n,v^n) v^n:\quad
\vert\vert v^n \vert\vert \leq K e^{\alpha t_n} \vert\vert v^0 \vert\vert
\quad \forall v^0,
$$
where $v$ the solution vector and $Q$ a matrix.
The stability criterion allows $e^{\alpha t_n}$ growth, but excludes $e^{\alpha n}$ growth,
{\it i.e.} for differential equations exponential growth is allowed in the continuum problem, 
but resolution dependent growth of the numerical algorithm is excluded.

To illustrate the analysis of numerical stability for a simple ODE problem, consider
the standard textbook example
$$
y' = \lambda y, \qquad y(0) = y_0
$$
and solve numerically with the forward Euler algorithm,
$$
y_{n+1} = y_n + h y_{n}' =  y_n + h \lambda y_n\ .
$$
The stability criterion yields $\vert y_{n+1} \vert / \vert y_n \vert = \vert 1 + h \lambda \vert$,
and the algorithm is unstable for $h > -2/\lambda$.
For $\lambda > 0$ the continuum solution grows exponentially and even stable algorithms 
will suffer from ill-conditioning.

\subsection{The wave equation as a toy model for hyperbolic equations}

The wave equation provides a standard example to illustrate well-posedness for 
evolution equations and to introduce different notions of the associated technical
concept of {\em hyperbolicity}. 
Starting with the second order form in 1D,
$\phi_{,tt} = c^2 \phi_{,xx}$,
we can obtain a mixed first/second order version,
$$
\phi_{,t} = c \pi, \quad \pi_{,t} = c \phi_{,xx}
$$
or a complete first order reduction
$$
\phi_{,t} = c \pi,       \quad \phi_{,x} = \psi, \quad
\pi_{,t}  = c \psi_{,x}, \quad  \psi_{,t} = c \pi_{,x},
$$
where $\phi_{,x} = \psi$ now plays the role of a constraint which is
preserved by the evolution equations:
$$
\partial_t (\phi_{,x} - \psi) = \partial_x \partial_t \phi - \partial_t \psi
= \partial_x \pi - \partial_x \pi = 0.
$$
Note that the evolution equation for $\phi$ decouples, and we may focus on
the system of equations for $\pi$ and $\psi$, which has the form
$$
\partial_t u = A \, \partial_x u, \quad u = \{\pi, \psi\}, \quad A =
\left(
\begin{array}{ll}
 0 & c \\
 c & 0
\end{array}
\right)\ .
$$
We call the matrix $A$, or generally the coefficients of the highest spatial
derivatives, the {\em principal part} of the system of partial differential
equations.
The characteristic variables $v_{\pm}=\psi \pm \pi$  correspond to the eigenvectors of $A$
and therefore satisfy advection equations, $\dot u_{\pm} = \pm c u_{,x}$,
where the eigenvalues of $A$ are $\pm c$.
We can easily construct the general solution in Fourier space:
$$
\Pi_\omega  = \Pi_0 \cos(c \, t \, \omega) + i \, \psi_0 \sin(c \, t \, \omega), \quad
\psi_\omega = \psi_0 \cos( c \, t \, \omega )  + i \, \Pi_0  \sin( c \, t \, \omega ).
$$
The norm $\vert \vert u \vert \vert^2 = \int \vert \Pi \vert ^2 + \vert \psi \vert^2$
can easily be checked to satisfy $\vert\vert u(t) \vert \vert = \vert \vert u(0) \vert \vert$, 
well-posedness can thus directly be read off from the general solution, and
numerical stability can be discussed in an analogous way.
General theorems can be used to show stability against perturbations with
lower order terms \cite{Gustafsson95}.

The system with principal part
\begin{equation}\label{1J-system}
A' =
\left(
\begin{array}{ll}
 1 & 1 \\
 0 & 1
\end{array}
\right),
\end{equation}
which we also will refer to as the 1-Jordan Block model,
shows very different features:
$A'$ has real eigenvalues -- $(1,1)$, but does not have a
complete set of eigenvectors (characteristics) and can not be diagonalized.
The solution is $U_\omega = (u, v)$,
$u=\omega t\sin\omega(t+ x) \, , \quad v=\sin\omega(t+ x)$.
Consider data with $u(0)=0$ and frequency $\omega$ -- in terms of the $L_2$
norm we now get
\begin{equation}
    \frac{\vert \vert U(t) \vert\vert}{\vert \vert U(0) \vert\vert} =
\sqrt{1 + t^2 \omega^2},
\end{equation}
which does not satisfy our criterion for well-posedness. It turns out, that
alternative norms can be chosen which render this problem well posed, but
generic lower order perturbations convert the frequency dependent linear growth
to frequency dependent exponential growth \cite{Gustafsson95}. The choice of
an appropriate norm is thus a subtle problem, and robustness with respect to
lower order terms a crucial criterion.

\subsection{First order systems}

Consider a linear first order system with constant coefficients:
$$
\dot u = A \partial_x u + B u + C,
$$
where $u$ is interpreted as a multi-component object. It is easy to see that for a single equation,
the function $A$ corresponds to the propagation speed of the wave. For a system of equations, the eigenvalues
of the matrix $A$ clearly can again be interpreted as propagation speeds if the matrix $A$ is diagonalizable.
The eigenvectors are then called the characteristic variables. 
General theorems allow to restrict analysis of well-posedness and stability to
$B = C = 0$ \cite{Gustafsson95}.
The formal solution in Fourier space is
$$
\dot{\hat u} = I \omega A \hat u \rightarrow \hat u = e^{I \omega A \,t} \hat u_0.
$$
In order to study whether the solutions can be bounded by initial data and thus whether the system
of equations allows a well-posed initial value problem, we need to evaluate  $e^{I \omega A}$.
Intuitively, no problems will arise if $A$ is diagonalizable with real eigenvalues, the solution
will then be purely oscillatory.

Well-posedness and stability can indeed be discussed in terms of eigenvalues
(characteristic speeds) and eigenvectors (characteristic variables) of $A$,
(again we refer to \cite{Gustafsson95} for an excellent textbook presentation):
\begin{itemize}
\item If all eigenvalues (speeds) are real, the system is called (weakly) hyperbolic,
and is well posed in absence of lower order terms in an
appropriate norm.
\item If a complete set of eigenvectors exists (the characteristic variables span the solution
  space), the system is called strongly hyperbolic, and admits a well posed initial value problem.
\item If the system is strongly hyperbolic and admits a conserved energy it is called symmetrizable
 (symmetric) hyperbolic (strongly hyperbolic implies symmetrizable in 1D).
\end{itemize}

As an example consider the York-ADM equations (\ref{def_K},\ref{eq:ADMconstraints},\ref{eq:ADMevol}) 
in 1D with lapse function $\alpha=\sqrt{\det g}$ (a standard choice),
linearized around Minkowski space:
$$
\begin{array}{l}
\dot h_{ii}=2 K_{ii}\ , \\
\dot K_{xx} = \frac{1}{2} \partial_{xx} h_{xx} + \partial_{xx}
  \left(h_{yy} + h_{zz} \right)\ , \\
\dot K_{jj} =  \frac{1}{2} \partial_{xx} h_{jj}  \qquad (j \, = \, y,z)\ . \\
\end{array}
$$
Computing the Jordan normal form of the matrix $A$ characterizing the principal part of the
first order system yields
$$
J(A) = \left(
\begin{array}{lllllll}
 0 & 0 & 0 & 0 & 0 & 0 & 0 \\
 0 & -1 & 0 & 0 & 0 & 0 & 0 \\
 0 & 0 & -1 & 1 & 0 & 0 & 0 \\
 0 & 0 & 0 & -1 & 0 & 0 & 0 \\
 0 & 0 & 0 & 0 & 1 & 0 & 0 \\
 0 & 0 & 0 & 0 & 0 & 1 & 1 \\
 0 & 0 & 0 & 0 & 0 & 0 & 1
\end{array}
\right).
$$
All characteristic speeds are real, but there are 2 Jordan blocks, the eigenvectors
of the characteristic matrix thus do not span the solution space and the system is
only weakly hyperbolic. Nevertheless, many physics results have been obtained
with the York-ADM system in spherical symmetry, and
excellent test results are obtained in the 1D test suites of \cite{Alcubierre2003:mexico-I-short}.
This is explained by noting that
decoupling the $xx$ components from the $yy$ and $zz$ components leads to
well--posed systems and good numerical results. This happens 
for the ``gauge wave'' metric eq. (\ref{eq:gaugewave}),
a linearized wave on flat background:
$$
  \label{eq:linearwave}
  ds^2=- dt^2 + dx^2 + (1+b) dy^2 + (1-b)dz^2, \quad
  b =  A \sin \left( \frac{2 \pi (x - t)}{d} \right),
$$
but also in spherical symmetry, where again the $yy$ and $zz$ components (or, say $\vartheta\vartheta$ and
 $\varphi\varphi$) of the metric are not independent.

\subsection{
Second order in space, first order in time systems}

First order in time, second order in space systems are typically used in astrophysically
oriented numerical relativity codes, because of their simplicity and a vague expectation
of better accuracy than first order reductions, which has at least partially been confirmed in
\cite{Calabrese:2005ft}.
Such systems are also attractive because they arise rather naturally in a Hamiltonian context.
The issue of well-posedness for such systems has however only been clarified very recently 
\cite{Nagy:2004td,Beyer:2004sv,Gundlach04a,Gundlach2004,Gundlach2006,Calabrese:2005ft},
for much of the history of NR it was widely believed that a clean route to well-posedness requires
the standard theory of first order hyperbolic systems, see {\it e.g.} 
\cite{Gustafsson95} for an excellent textbook reference.

Again, the wave equation  provides an excellent simple example of what is going on.
The first order in time, second order in space wave equation is
$$\partial_t \phi(t, x) = \Pi(t, x), \quad
\partial_t \Pi(t, x) = \partial_{xx} \phi(t, x).
$$
Consider the family of solutions \cite{Frittelli:2000uj}
$$
\phi(x, t) = \sin(\omega x) \cos(\omega t),\qquad
\Pi(x, t) = -\omega \sin(\omega x) \sin(\omega t)
$$
with initial data $\phi_0 (x) = \sin(\omega x), \Pi_0( x) =0$.
Varying $\omega$ in the initial data, the $L_2$ norm of the solution at
time $t$,
$\int_{0}^{2\pi} (\vert \phi\vert^2 + \vert \Pi\vert^2) \, dx$,
can be made arbitrarily large with respect to the initial data (whose norm is
independent of $\omega$), this
contradicts well-posedness in $L_2$. Since numerical codes implementing these ill-posed equations,
as well as the York-ADM evolution equations (\ref{eq:ADMevol}) which had been argued to lead to an ill-posed 
initial value problem before, did not show blowup, it was suggested  \cite{Frittelli:2000uj} 
that well-posedness may not be crucial for numerical relativity codes.

The issue is resolved as follows: since all the solutions for the wave equation are actually 
bounded and show no growth, ill-posedness in  $L_2$ is not a sign of pathology in this case --  the problem rather 
is the inappropriate use of the $L_2$-norm.
Translating the first order norm that one gets from introducing
$$
X = \partial_x  \phi,
$$
which leads to a well posed, symmetric hyperbolic problem, 
to the second order variables, and using the norm implied by this translation
$$
\int_{0}^{2\pi} (\vert \Pi \vert^2 + \vert \partial_x \phi\vert^2) \, dx
$$
does in fact yield well-posedness, as can be shown by pseudo-differential
reduction ({\it i.e.} solving
explicitly in Fourier-space) \cite{Nagy:2004td}.
Note that on the discrete level an ambiguity regarding the
discrete norm needs to be resolved: it is not clear how the derivative should be discretized, 
or whether this matters at all. This ambiguity is resolved in \cite{Calabrese:2005ft}, see below.
The reason that a code that implements the first order in time, second order in space wave equation
does not show any pathological behavior is thus simple: the system is well posed, and as shown
in \cite{Calabrese:2005ft} can be discretized stably in a straightforward way. The York-ADM equations are however indeed
only weakly hyperbolic and therefore lead to an ill-posed initial value problem. However, the growth is 
only linear, and is easily dismissed without careful convergence tests, or when artificial dissipation is 
added. Using the norm that would correspond to a first order-reduction of the system notions of well-posedness
and hyperbolicity can then indeed be defined for mixed order hyperbolic systems 
\cite{Nagy:2004td,Beyer:2004sv,Gundlach04a,Gundlach2004,Calabrese:2005ft}.

A simple  ``normal form'' for mixed order hyperbolic systems has been introduced in \cite{Calabrese:2005ft}:
$$
\p_t \bm{u} = P \bm u\,,\quad
\bm{u} = \columntwo u v, \quad  P =
\matrixtwotwo{A^i \p_i + B}                 {C}
             {D^{ij} \p_i \p_j + E^i\p_i+F} {G^i\p_i+J}.
$$
Evolved variables are split into two types: $u$ are differentiated twice and
$v$ are not. Not all second order in space systems
can be written in this form (a simple  example is the heat equation, $u_t = u_{xx}$, which is parabolic).
Now consider the {\em second order principal
symbol $\hat P'$} and a matrix $E$, which is the principal
part of the associated first order system:
$$
  \hat P' =
  \matrixtwotwo    {i \omega A^n}      {C}
                   {- \omega^2 D^{nn}} {i \omega G^n}, \quad
  \hat E = i \omega
  \matrixthreethree {0} {0}      {0}
                    {0} {A^n}    {C}
                    {0} {D^{nn}} {G^n}.
$$
The main result on the continuum level of the equations is that if
$E$ is diagonalizable in a regular way, the initial value problem is well posed:
$$
\|\bm{u}(t,\cdot)\| \le K e^{\alpha t} \|\bm{u}(0,\cdot)\|, \quad
\|\bm{u}\|^2 \equiv \int |u|^2 + \sum_{i=1}^d|\p_i u|^2 + |v|^2 d^dx\, ,
$$
where the norm is again the straightforward translation of the $L^2$ norm of the associated
first order system. This result is equivalent to results that had previously
been obtained by other authors
in \cite{Nagy:2004td,Gundlach04a,Gundlach:2005ta}.
The result may seem trivial, but has ended a long controversy in numerical relativity about whether
the evolution systems typically used in mainstream numerical relativity are well posed, or even
whether their well-posedness can be discussed in a mathematically rigorous way. 
Recent results along the lines sketched here prove well-posedness results for
the evolution systems and coordinate gauge conditions that are actually used in large
scale binary black hole evolutions \cite{Gundlach:2006tw}, which had 
seemed far out of reach only a few years ago.

The appropriate norm for the discrete case is
$$
\| . \|  \rightarrow \|\bm{u}\|_{h,D_+}^2 \equiv \|u\|_h^2 + \|v\|_h^2 +
\sum_{i=1}^d\|D_{\pm i} u\|_h^2.
$$
The ambiguity of discretizing the derivative in the norm is resolved by using the
first order forward ($D_+$) or backward ($D_-$) spatial derivative in combination
with standard centered second or fourth order differencing. Using centered derivatives
in the norm is shown not to be robust with regard to perturbations in lower order terms.

Note that for the continuum version of the matrix $E$ the frequency $\omega$ can be  factored out,
which is not generally the case for the discrete version of $E$, this would only hold in general if
the second derivative is really also the derivative of the first derivative on the discrete level.
For some evolution systems in general relativity certain standard discretizations are thus problematic,
as discussed in \cite{Calabrese:2005ft}. 

\subsection{Stable and well-posed is not enough}

Clearly, well-posedness and numerical stability are not sufficient to guarantee successful numerical
simulations, since exponential growth or blowup in finite time, which is allowed for well posed problems,
is typically not tolerable for numerics (unless the timescale of the growth is sufficiently smaller than
the timescale of interest in the solution).
A typical trick to get rid of exponential growth is a change of variables, but this is in general difficult for
tensors.
As an example, it is a standard result of textbooks on ordinary differential equations that
sufficiently regular ODE initial value problems are always well--posed,
solutions may however blow up in finite time:
$$
y' = \lambda y^2, y(0) = y_0 \quad \rightarrow \quad
y(x) = y_0/(x \, y_0 - 1).
$$

In numerical relativity, exponential growth of the continuum solution
already appears in seemingly trivial test problems.
Consider a ``gauge wave'' metric -- the flat metric of Minkowski space
in coordinates that correspond to a traveling wave of coordinate distortion.
This metric provides a challenging test for most evolution codes and is part of a numerical relativity
test suite \cite{Alcubierre2003:mexico-I-short}:
\begin{equation}\label{eq:gaugewave}
  ds^2=-H dt^2 +Hdx^2+dy^2+dz^2,\quad
  H = H(x-t)=1 - A \sin \left( \frac{2 \pi (x - t)}{d} \right).
\end{equation}
This line element represents Minkowski spacetime with a nontrivial
choice of harmonic time slicing ($\nabla^a \nabla_a t = 0$). 
One analytic source of rapidly growing
error is the instability of flat space on $T^3$. Another problem is the
existence of a family of harmonic, exponential gauge modes corresponding to $H
\rightarrow e^{\lambda t} H$, with arbitrary $\lambda$.
This problem can be modeled by
a nonlinear wave propagating on Minkowski space \cite{Babiuc:2004pi}:
$$
\eta^{ab}\partial_a \partial_b \Phi - \frac{1}{\Phi}\eta^{ab}(\partial_a \Phi)
(\partial_b \Phi) = 0 = \Phi \eta^{ab}\partial_a \partial_b \log\Phi.
$$
Imposing periodic boundary conditions, one evolves on a 3-torus, {\it i.e.} there
are no boundaries.
For $\Phi > 0$ the Cauchy problem is well-posed. In addition to
the solution $\Phi = 1 + F(t-z)$, $F>-1$, the system also admits
the solutions
$$
\Phi_\lambda = e^{\lambda t} \, (1 + F(t-z))
$$
for arbitrary $\lambda$.
Numerical errors will excite the growing modes, which eventually dominate the
signal.
As pointed out in \cite{Babiuc:2004pi}, an obvious solution is to use $\log\Phi$ as an evolution variable,
but the example just models a situation that does arise in numerical relativity, where it is not
clear how to take the logarithm of the metric.

For a problem that is actually ill--posed, the situation will be much worse
than for our gauge-wave toy problem: 
typically higher frequencies will correspond to faster growth ({\it i.e.} larger
$a, K$ in the estimate (\ref{wellposed})). Since better resolution of the grid
does in particular allow higher frequencies, improving the resolution will in general
lead to a numerical solution that grows faster.
This is analogous to an unstable numerical scheme -- in such a case, the 
discretized equations do not lead to a  well posed problem.

\section{Energy, momentum, and radiation}\label{sec:energy}

For astrophysical systems like stars or inspiraling black holes, 
it is physically reasonable to assume that spacetime  becomes flat
in the limit of large distance, and that the systems can be considered
``isolated'', {\it i.e.}
unaffected by the large scale structure of the universe.
The formulation of a rigorous concept of ``asymptotic flatness'' in GR is far from
straightforward, due to the absence of a background metric or
preferred coordinate system, in terms of which falloff rates can be
specified. There are two possible routes to overcome this problem:
In the first approach one simply assumes the existence of a suitable
coordinate system, which is then used to formulate falloff conditions
for tensor components in this coordinate system. We will follow this approach below
to define quantities such as the total energy of a gravitating system. The obvious drawback is
that taking limits is often problematic, in particular in a numerical
context, and coordinate invariance of expressions has to
be carefully checked. 
A resolution of these problems is provided by a definition of asymptotic
flatness, where, after a suitable conformal rescaling of the metric, ``points
at infinity'' are added to the manifold, one thus works on a compactified
auxiliary manifold, and local differential geometry can be used to study
the asymptotic properties of the gravitational field \cite{Penrose1963}.
Note that the notion of asymptotic flatness at timelike infinity does not make
much sense in a general situation, because then all energy would have to be
radiated away, leaving only flat space behind -- excluding black holes or
``stars''. The important notions are asymptotic flatness in spacelike and null (lightlike) directions.
A detailed discussion of the global structure of spacetimes describing isolated systems 
in general relativity is not really possible here -- an excellent overview 
is given in the textbook  \cite{Wald84}, for work in the context of numerical relativity
see {\it e.g.} \cite{Winicour98,Husa01,Andersson:2002gn,Frauendiener04,Husa:2005ns}.

The concept of asymptotic flatness of isolated systems is intimately related
to the possibility of defining the total energy-momentum for such systems
in general relativity.
In GR there exists no known well-defined local energy density of the
gravitational field, but a total energy-momentum, which transforms
as a 4-vector under asymptotic Lorentz transformations, can be
assigned to an isolated system \cite{Wald84}, analogous to
a particle in special relativity. It is a constant of
motion and can therefore be expressed in terms of the initial data on an
asymptotically flat Cauchy hypersurface.

If a manifold has more than one asymptotically flat end, {\it e.g.} in the
presence of wormholes of the Einstein-Rosen-bridge type, then -- in general
different -- masses can be associated with each of these asymptotic regions.

The expression for the energy momentum of GR at spatial infinity
has been given first by Arnowitt, Deser and Misner in 1962 \cite{Arnowitt62}
in the context of the Hamiltonian formalism, and is usually called the ADM
momentum, the time component being called the ADM energy or, somewhat
inconsistently, the ADM mass, different from the rest mass to be defined
below.

The expressions  for the mass and momentum are given as limits of surface integrals over non-covariant
quantities, and have to be evaluated in asymptotically Cartesian (regular)
coordinates $\{x^{i}\}$ -- where the components of the metric tend to
diag$(1,1,1)$ for large radii $r=\sqrt{x_{i}x^{i}}$. The surfaces are spheres
$S_{r}$ of radius $r$.

We define the surface integrals (which we will also refer to as ADM integrals)
\begin{eqnarray}
E(r) & = & \frac{1}{16\pi}
\int_{S_{r}}\sqrt{g}g^{ij}g^{kl}\left(g_{ik,j}-g_{ij,k}\right)dS_{l},
\label{madm_int} \\
P_{j}(r) & = & \frac{1}{8\pi}\int_{S_{r}}\sqrt{g}\left (K^{i}_{j} -
  \delta^{i}_{j}K \right)dS_{i}, \label{padm_int} \\
J_{j}(r) & = & \frac{1}{8\pi} \epsilon_{jl}{}^m \int_{S_{r}} \sqrt{g} x^l
\left(K^i_m - K \delta^i_m \right) dS_{i} \label{jadm_int}
\end{eqnarray}
which  have to be evaluated in an asymptotically Cartesian coordinate system.

The ADM energy $M_{ADM}$ and linear and angular momentum $P_j$ and $J_j$ are
then given by \cite{Omurchadha74,York79}
\begin{equation}\label{madm}
M_{ADM} =\lim_{r\rightarrow\infty} E(r),\quad
P_{j} = \lim_{r\rightarrow\infty} P_{j}(r), \quad
J_{j} = \lim_{r\rightarrow\infty} J_{j}(r),
\end{equation}
and the rest mass $M_{R}$ can be defined as $M_{R}^2=M_{ADM}^{2}-\sum_{j=1,3}P_{j}P_{j}$.

For an asymptotically Schwarzschildian metric $h_{ab}$ (here the extrinsic curvature
falls off faster than in the general case , which allows for a boost, {\it i.e.} linear momentum) 
the ADM mass can be read off directly as the $\frac{1}{r}$-term of the metric:
\begin{equation}\label{mono}
h_{ab}=\left(1+\frac{m}{2r}\right)^{4}\delta_{ab}+q_{ab},\qquad \mid q_{ab}
\mid\leq\frac{const.}{1+r^{2}}.
\end{equation}

A fundamental issue of GR is the positivity of the ADM energy, since if
the energy of an isolated system can be negative, it would most likely be
unstable and decay to lower and lower energies.
While it is trivial to write down a metric with negative mass,
for reasonable matter fields with nonnegative energy density (thus
satisfying the dominant energy condition), nonnegativity of the ADM energy
thus can be expected on physical grounds. Indeed a complete proof
of this positive energy conjecture has been given in 1982 by
Schoen \& Yau \cite{Schoen79b} (several simplified proofs have been given
afterwards).

For radiation processes we also require definitions of total energy, linear
and angular momentum that decrease as energy and linear as well as angular
momentum are radiated to infinity. The appropriate quantities are the Bondi
quantities \cite{Bondi62}, which can be defined as taking the limit of the ADM
integrals not toward spatial infinity, but rather toward null infinity
\cite{Katz88,Brown:1996bw,Poisson04a}, {\it i.e.}, the limit to infinite distance is
taken for constant retarded time instead of on a fixed Cauchy slice. In the
context of our numerical treatment, the ADM and Bondi quantities can be calculated
by computing values at several radii, and then performing a
Richardson extrapolation (in extraction radius, not, as is more usual, in grid spacing). 
Here the Bondi quantities can be computed at
any time for a fixed extraction radius, and have to be compared between
different radii by taking into account the light travel time between the
timelike cylinders of different radii, see {\it e.g.} \cite{Bruegmann:2006at}.

Radiation quantities are conveniently defined in terms of the (complex) Bondi news function
${\cal N}(t) := \partial_t (h_+ - I h_\times)$, which is the time derivative of the complex
strain $h$ taken at null infinity, where $h_+$ and $h_\times$ are the two polarization modes
of the gravitational  field, see eq. (\ref{eq:hpluscross}). 
The expressions for the radiated energy and momenta then become
\begin{eqnarray}
  \frac{dE}{dt} &=& \frac{1}{16\pi} \int_{\Omega} {\cal N}\overline{\cal N} d{\Omega}, \\
  \frac{dP_i}{dt} &=& - \frac{1}{16\pi}
      \int_{\Omega} \ell_i {\cal N}\overline{\cal N} d{\Omega}, \\
  \frac{dJ_z}{dt} &=& - \frac{1}{16\pi}
      \mathrm{Re}\left[ \int_{\Omega} {\cal N}_{,\phi}
      \left( \int_{-\infty}^t \overline{{\cal N}} d\hat{t} \right)d{\Omega} \right],
\end{eqnarray}
where
$$
  \ell_i = \left(-\sin \theta \cos \phi,\,\,\,-\sin \theta \sin \phi,\,\,\,
           -\cos \theta \right),
$$
and an overbar denotes complex conjugation.
The strain $h$ is most often computed indirectly via double time integration of
 certain projections of the curvature tensor, 
see {\it e.g.} \cite{Bruegmann:2006at}
for a detailed recent description in the context of numerical relativity.

\section{Horizons}\label{sec:horizon}

This section introduces the concept of trapped surfaces and
apparent horizons.
These are intimately related to two of the most fascinating features of
general relativity: the appearance of singularities and the appearance of
causal membranes: event horizons that enclose so-called black holes -- regions
of spacetime that do not allow any information to escape to the outside world.

For quite some time it was not clear whether singularities, 
which have first been
found in highly symmetrical spacetimes such as the
Friedmann-Robertson-Walker cosmologies or the Schwarzschild spacetime,
are really generic, or mere artifacts of the high symmetry in these situations.
Only in 1965 the generic character of singularities has been proven in a
theorem
by Penrose \cite{Penrose65}. A variety of similar theorems with modified assumptions
has followed, for an overview see {\it e.g.} Hawking and Ellis \cite{Hawking73a}
or Wald \cite{Wald84}.

A crucial notion in the singularity theorems is that of a {\em trapped surface}:
a spacelike 2-surface with the property that the area of
the outgoing wavefront {\em decreases} toward the future.
The central idea of the
singularity theorems is then, that provided some energy condition ({\it e.g.} that
the local energy density of matter is nonnegative) holds for
the matter fields, gravity is always attractive, and light rays always get
focused.
Thus, if a region of spacetime starts to collapse, as is signaled by the
appearance of a trapped surface, this collapse cannot be halted and will
continue until a spacetime singularity forms.

The original Penrose singularity theorem \cite{Penrose65} states that, provided
a closed
trapped surface exists and the Cauchy surface is non-compact, at least one of
 the following things will happen:
\begin{enumerate}
\item
There occurs negative local energy density.
\item
Einstein's equations are violated.
\item
The spacetime manifold is incomplete -- a singularity occurs.
\item
The concept of spacetime loses its meaning at very high curvatures,
possibly due to quantum gravity effects.
\end{enumerate}
This means, that in the framework of classical general relativity
a trapped surface signals the occurrence of a singularity in the future.

While the singularity theorems have shown that sufficiently ``strong''
initial data will develop a singularity, these
theorems make no statement about the nature of the singularities. It is of
particular interest, whether the singularities that arise in a physically
reasonable collapse situation is visible to any observer.

A wide-spread hope -- at least in classical general relativity -- is
expressed by the cosmic-censorship hypothesis of Penrose
\cite{Penrose69}, which is by now floating through the literature in various
formulations. The basic idea is that naked singularities -- that is
singularities that can be seen by outside observers -- should not
arise from regular initial data. The future light cone of the beginning of
a naked
singularity would be a Cauchy horizon and destroy predictability. A
distinction is made between global and
local nakedness. Globally naked singularities can influence asymptotic
infinity  (in an asymptotically flat setting) -- they are not hidden by an
event horizon. Inside of an event horizon, there may sit a locally naked
singularity -- while no information can escape from it to infinity it can
still be seen by observers inside of the black hole. To rule out locally
naked singularities all singularities have to be spacelike, which is equivalent
to the spacetime being globally hyperbolic. In the weaker formulation,
global hyperbolicity is only required outside of an event horizon.

Counter-examples to overly restrictive formulations of cosmic-censorship are
known in various cases, {\it e.g.} for phenomenological matter like dust outside of
the physical validity of the equation of state, but also for sets of
measure zero in the space of initial data (related to a collection of phenomena
known as ``critical collapse'', see {\it e.g.} \cite{Gundlach00a} for an extensive overview).

Black holes, although dangerous and exotic, from
this viewpoint are the {\em good guys}, the {\em bad guys} are naked
singularities.
From the viewpoint of quantum gravity, a violation of cosmic censorship
actually seems rather desirable, since then regions of strong curvature,
where quantum effects are important, may actually be observable in the
outside world.

A notion that is derived from that of a trapped surface, is the {\em apparent
horizon}: the boundary of the region of trapped surfaces, which turns out to be ``marginally
trapped'' -- {\it i.e.} the area of
the outgoing wavefront neither increases nor decreases toward the future. 
In contrast to event horizons, which are defined in a global manner and
can only be determined if the maximal time development is known, apparent
horizons and trapped surfaces are defined locally in time, within a single
slice, and can be directly determined from the initial data.
Since the event horizon can only be defined if a suitable notion of infinity
exists in a given spacetime, one could take the point of view of defining a
black hole directly by means of apparent horizons \cite{Ashtekar:2004cn}.
The location in a given geometry and the analysis of the physical properties of the apparent horizon 
thus play an important role in numerical relativity. ``Apparent horizon finders'' are an essential
component of all binary black hole evolution codes and discussed extensively in \cite{Thornburg:2006zb}.

\section{Initial data for numerical relativity}\label{sec:id}

\subsection{The conformal approach to solve the constraints}

We have already taken a first step  in identifying freely
specifiable data for Einstein's equations by formulating them as a Cauchy
problem. Given appropriate initial data on a spacelike hypersurface,
the spacetime is determined as a unique time development (modulo
diffeomorphisms) of these data
for future and past times. Due to the presence of the constraint equations
(\ref{eq:ADMconstraints})
the initial data can not be specified freely, and the next task therefore is to
extract the unconstrained part of the initial data in such a way that the
constraints then uniquely determine the whole set of initial data, and thus
the whole spacetime.
There exists a standard formalism to accomplish this goal and
solve the constraints, the conformal approach developed by Lichnerowicz, York,
\'O Murchadha and others \cite{Choquet80}.
The conformal approach is based on conformal rescaling, in particular of the
spatial metric $h_{ab}$, which is expressed from a 'base metric'
$\bar h_{ab}$ and a conformal factor $\psi$ as
\begin{equation}\label{def_tilde_h}
h_{ab} = \psi^{4}\bar h_{ab},\qquad\psi>0.
\end{equation}
By combination of (\ref{def_tilde_h}) with
a similar rescaling and decomposition of the extrinsic curvature into a
traceless symmetric part and a part that is derived from a vector potential,
the constraints will be written as a coupled set of four well posed
quasilinear elliptic PDEs for four 'gravitational potentials'.

For the conformal rescaling (\ref{def_tilde_h}) the scalar curvature
transforms as
$$
R_{h}=\psi^{-5}\left(R_{\bar h}\psi-8\lap_{\bar h} \psi\right).
$$
Next we define the trace-free part of the extrinsic curvature by
$$
A^{ab}:= K^{ab}-\frac{1}{3}h^{ab}K.
$$
The rescaling
\begin{equation}\label{scale_Aupper}
A^{ab}=\psi^{10}\bar A^{ab}
\end{equation}
then results in the property
\begin{equation}\label{A^ab rescaled}
D_{a}A^{ab} = \psi^{-10}\bar D_{a}\bar A^{ab},
\end{equation}
where $\bar D_{a}$ is the unique derivative operator associated with the
metric $\bar h_{ab}$. For $K=const.$, eq. (\ref{A^ab rescaled}) expresses the
fact, that the momentum constraint is conformally invariant, if the
tracefree part of the extrinsic curvature is properly rescaled.
Note that the definition of $A^{ab}$ in eq. (\ref{scale_Aupper}) is
equivalent to setting
$$
A_{ab}=\psi^{2}\bar A_{ab},
$$
where
$$
A_{ab}= A^{cd}h_{ac}h_{bd},\qquad \bar A_{ab}= \bar A^{cd}
\bar h_{ac}\bar h_{bd}.
$$
The traceless symmetric tensor $\bar A^{ab}$ is decomposed as
$$
\bar A^{ab} = \bar A^{ab}_{TT} + (LW)^{ab}
$$
into a
divergence-free (transverse) traceless part $\bar A^{ab}_{TT}$ and a
tracefree part that can be obtained from a potential $W^{a}$,
$$
(LW)^{ab}:=\bar D^{a}W^{b} + \bar D^{b}W^{a} - \frac{2}{3}\bar h^{ab}
 \bar D^{c}W^{c}.
$$
Insertion of the reverse decomposition
$$
\bar A^{ab}_{TT} = \bar A^{ab} - (LW)^{ab}
$$
into the constraint equations (\ref{eq:ADMconstraints}) yields
\begin{eqnarray}
 \label{MC_conformal}
D_{a}(LW)^{ab}=-\bar D_{a}\bar A^{ab} + \frac{2}{3}\psi^{6}D^{b}K
  + 8\pi\psi^{10}j^{b},\\
\label{HC_conformal}
-\bigtriangleup_{\bar{h}}\psi+\frac{1}{8}R_{\bar h}\psi
-\frac{1}{8}\psi^{-7}(\bar A^{ab} - (LW)^{ab})^{2}
+\frac{1}{12}\psi^{5}K^{2} & = & 2\pi\psi^{5}\rho,
\end{eqnarray}
where we have now kept the energy density  $\rho$ and momentum density $j^{b}$
of matter fields in the expressions.
The freely specifiable quantities here are the metric $\bar h_{ab}$,
the trace of the extrinsic curvature, $K$, and a symmetric tracefree tensor
$\bar A^{ab}$, which together comprise the local freedom in choosing
initial data. The constraint equations now take the form of a coupled elliptic
system of PDEs for the 'potentials' $\psi$ and $W^{a}$, the initial data are
reconstructed as
$$
h_{ab} = \psi^{4}\bar h_{ab}, \qquad
K^{ab} = (\bar
A^{ab}-(LW)^{ab})^{2}\psi^{-10}+\frac{1}{3}K\psi^{-4}h^{ab}.
$$

Prior to the choice of a set of fields $(\bar h_{ab},K,\bar A^{ab})$,
one  has to specify a 3-manifold $S$, on which the fields
$(\bar h_{ab},K,\bar A^{ab})$ are defined and the equations
(\ref{HC_conformal},\ref{MC_conformal}) are to be solved. If the manifold
$S$ has a nonempty boundary, or is not compact, it will be necessary to
impose boundary conditions or asymptotic falloff conditions, which have to
be chosen, along with the topology of $S$, on physical grounds.

Due to the diffeomorphism invariance of GR, different initial data sets
will give rise to the same spacetime ({\it e.g.} different slices of the same
spacetime in different coordinate systems), which leads to the question for
 the
number of local {\em physical} degrees of freedom represented
by the initial data  $(\bar h_{ab},K,\bar A^{ab})$.
Since  $(\bar h_{ab}$ and $K^{ab})$ are both symmetric, the initial data
are represented by 12 free functions. Three of these correspond
to conditions on the spatial coordinate system (the metric $\bar h_{ab}$
can be regarded as given by three functions by choosing a coordinate system
where it is diagonal). In addition to initial data which  are equivalent
with respect to spatial diffeomorphisms, also data on different Cauchy
surfaces may give rise to the same spacetime.
A hypersurface is specified by one function, which can usefully be identified
with the trace of the extrinsic curvature, also called the mean curvature --
since it corresponds to an average over the components of the extrinsic
curvature.
By imposing the 4 constraint equations on the remaining 8 free functions and
dividing by two, one arrives at 2 local degrees of freedom per space point
(confirming that GR is indeed a field theory).
This is the same number of degrees of freedom as results from the linearized
theory, which is the theory of a spin 2 field, where the 2 degrees of
freedom can be identified with 2 independent states of polarization --
just as in the case of electromagnetism (see {\it e.g.} ref. \cite{Wald84}, sec.
4.4).

Most discussions of initial data restrict attention to so-called constant
mean curvature (CMC) hypersurfaces. The condition that the mean curvature, or
equivalently the trace of the extrinsic curvature, be constant is a
coordinate independent statement and decouples the system
(\ref{HC_conformal},\ref{MC_conformal}). The procedure of solution then
becomes the following:
\begin{enumerate}
\item
Choose  $\bar h_{ab},\bar A^{ab}, K=const.$,
\item
solve (\ref{MC_conformal}) for $K^{ab}$,
\item
solve (\ref{HC_conformal}) for the scale factor $\psi$, regarding
the extrinsic curvature term as source.
\end{enumerate}
Slices with $K=0$ are called maximal
slices. Maximal hypersurfaces embedded in a Lorentzian manifold locally maximize the
3-dimensional volume in the same way as a timelike geodesic maximizes
the proper length, or a minimal 2-surface in Riemannian 3-space minimizes
the area.
The opposite extreme, {\it i.e.} a minimal 3-slice, or a maximal 2-surface is
not possible, since deformations of the submanifold that just make the
embedding  'more wiggly' will result in a change of volume of a definite
sign: it is positive for a Riemannian hypersurface or a timelike curve
in a Lorentzian manifold, and negative for a submanifold of a Riemannian
manifold.
The concept of extremal submanifolds thus is a natural generalization
of geodesics as straightest curves.

Considering the fact, that even simple metrics can be made to look arbitrarily
complicated by coordinate choice, it is natural to specify initial data
on hypersurfaces that are embedded as simply as possible, which leads to
the consideration of maximal (ore more general CMC) slices as 'least wiggly'.
Another advantage of maximal slices is that a foliation of maximal slices
avoids singularities \cite{Estabrook73}, which has been utilized in many numerical
calculations.

Initial data for black
holes can conveniently be constructed by ``filling'' the spacetime volume
inside the horizon with artificial asymptotically flat ends.
This construction enforces the presence of ``throats'' -- minimal surfaces, and for 
non-vanishing extrinsic curvature also the presence of horizons (for
vanishing extrinsic curvature the outermost minimal surfaces is actually an apparent horizon). 
These
asymptotic regions are typically compactified for technical convenience,
rendering the nontrivial topology of the resulting spacetime representable
on $R^3$, or $S^3$ if the physical asymptotically flat end is also compactified.
Compactification naturally leads to singular behavior at the coordinate locations of
the artificial asymptotic ends, which are commonly referred to as ``punctures''.
The treatment of the singularity in the constraint equations is well understood
\cite{Dain:2002vm}. The use of such initial data has first been advocated in \cite{Beig94},
and has become the method of choice for a large fraction of work on
binary black holes in numerical relativity following the prescription of Brandt
and Br\"ugmann \cite{Brandt97b}.
The standard simplifying assumptions in the binary black hole literature are that
the spatial geometry is conformally flat and that
the extrinsic curvature is of the Bowen-York form \cite{Bowen80}, which is a family of solutions
to the momentum constraint on a flat background, for which the total linear and angular 
momentum can be freely specified. By linearity of the momentum constraint and the flat
background, superposition can be applied to construct multiple--black hole solutions.
Removing the assumption of conformal flatness becomes an issue particularly for spinning black holes,
see \cite{Hannam:2006zt} for a recent discussion.

An essential question regarding the construction of initial data is 
the issue of physical validity of the initial data set, say for the inspiral of two black holes.
Data sets should contain little or no ``artifical radiation'', and should correspond to the
actual inspiral of astrophysical compact objects. The typical aim is
to start with initial data that correspond to the astrophysically most
relevant case of quasi-circular inspiral, which essentially means that
the orbital eccentricity is very small: eccentricity is radiated away rather efficiently for
inspiraling black holes \cite{Peters:1964}. It has recently been possible for the nonspinning equal mass
case to directly use
the initial momenta (and separations) computed from a PN inspiral, and show that
not only the eccentricity of such initial data is very small, but also that the influence
of ``artificial radiation'' inherent in the initial data can essentially be neglected.
I expect this method to carry over to more general scenarios, for large spins of the black holes
non-conformally flat initial data might have to be used along the lines of \cite{Hannam:2006zt}
and the references cited therein. For extensive reviews of the problem of construction initial 
data for general relativity see \cite{Cook00a,Gourgoulhon:2007ue}.

\section{The binary black hole revolution}\label{sec:results}

The inspiral and collision of compact objects -- in particular of black hole binary systems --
is considered one of the most important sources of gravitational waves for earth- and space-based
detectors. Producing templates for gravitational-wave data analysis that describe signals from inspiraling
compact binaries will require large parameter studies, and correspondingly large computational
resources: The eventual goal of our simulations is to map the physical parameter space of
gravitational wave signals from black hole coalescence, which is
essentially given by the mass ratio and individual spins, as well as the initial
orientation of the spins. The latter determines in particular the spin orientation at
merger time, which may have a significant influence on the gravitational wave signal.

In order to produce ``complete'' waveforms, which contain large numbers
of gravitational wave cycles from the inspiral phase, as well as the merger
and ringdown phases, it is necessary to start the numerical simulations in the regime where
Post-Newtonian analytical calculations are valid. These describe very accurately
the waveforms of the early inspiral process, but break down for small separations of the
black holes. This ``matching'' of analytical and numerical results requires large
initial black-hole separations and large integration times.

The numerical solution of the full Einstein equations represents a
very complex problem, and for two black holes the spacetime
singularities that are encountered in the interior of black holes pose
an additional challenge.
In order to obtain accurate results both the
use of mesh refinement techniques {\em and} a good choice of
coordinate gauge are essential.  Together with the complicated
structure of the equations --- a typical code has between ten and
several dozen evolution variables, and, when expanded, the right hand
sides of the evolution equations have thousands of terms --- this
yields a computationally very complex and mathematically very subtle
problem.

For a long time, typical runs had been severely limited by the
achievable evolution time before the simulations became too inaccurate
or before the computer code became unstable, and there were serious
doubts whether numerical relativity techniques could produce
gravitational-wave templates, at least in the near future.  This
picture has drastically changed ever since in spring of 2005
Pretorius~\cite{Pretorius:2005gq} presented the first simulation
lasting for several orbits, using adaptive mesh refinement,
second-order finite differencing, a sophisticated method to excise the
singular interior of the black hole from the grid, and an implicit
evolution algorithm.

An alternative to the ``excision'' method of treating black holes
is to ``fill'' the black hole with a
topological defect in the form of an interior space-like asymptotic
end, the ``puncture'' \cite{Brandt97b}, and to freeze
the evolution of the asymptotic region through a judicious choice of
coordinate
gauge~\cite{Bruegmann97,Alcubierre00a,Alcubierre02a}.
The latter approach, combined with a setup where the topological
defect is allowed to move across the grid (``moving puncture''
approach~\cite{Campanelli:2005dd,Baker05a}) has led to a giant leap
forward in the field
~\cite{Herrmann:2006ks,Campanelli:2006gf,Baker:2006yw,Baker:2006nr,Campanelli:2006uy,Sperhake:2006cy,Hannam:2006vv},
taking the first orbit simulations of black
holes~\cite{Bruegmann:2003aw,Pretorius:2005gq} to more than ten
orbits and allowing accurate wave extraction.

\begin{figure}[t!]
\begin{center}
\includegraphics[width=6cm]{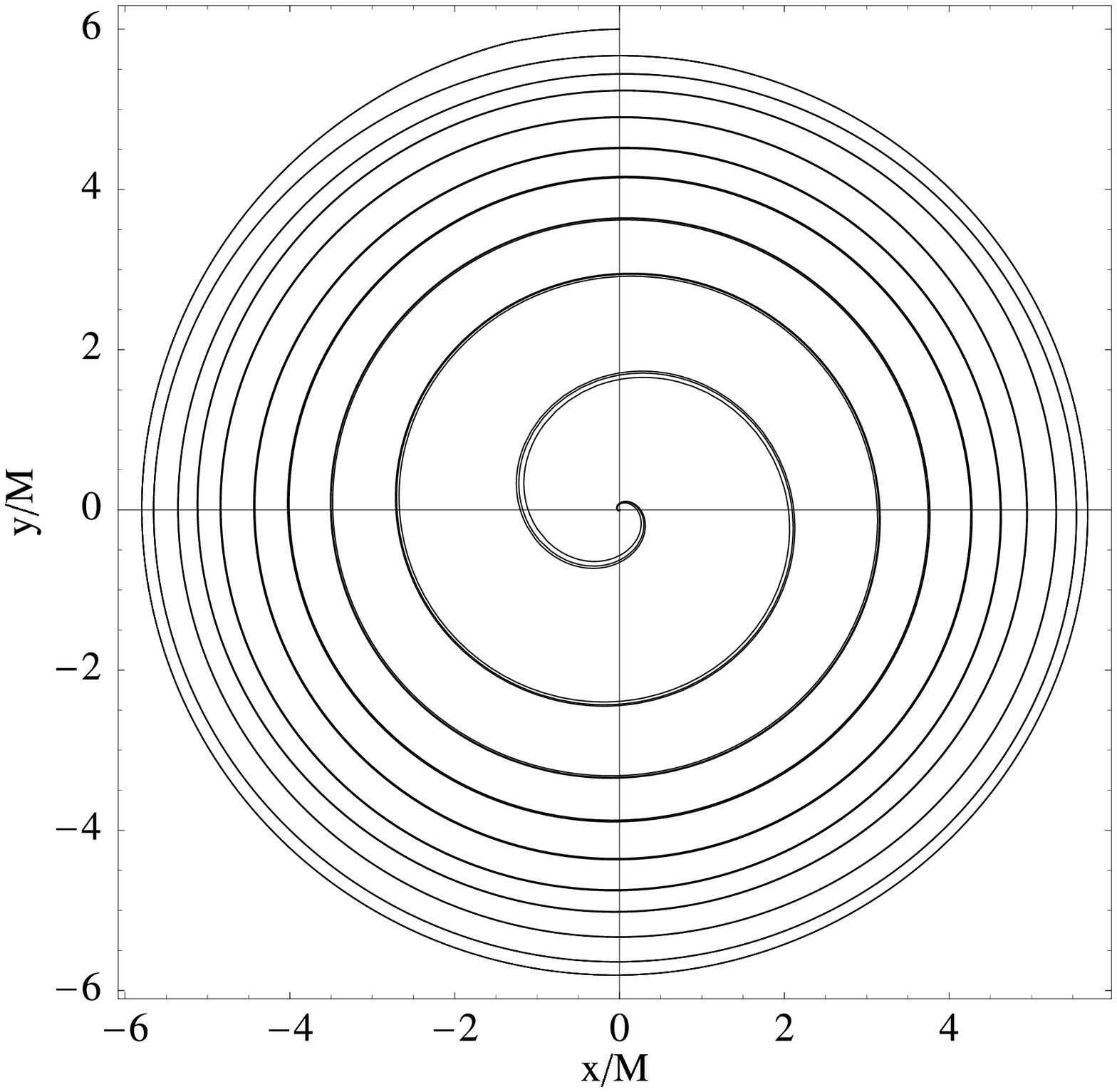}
\includegraphics[width=6cm]{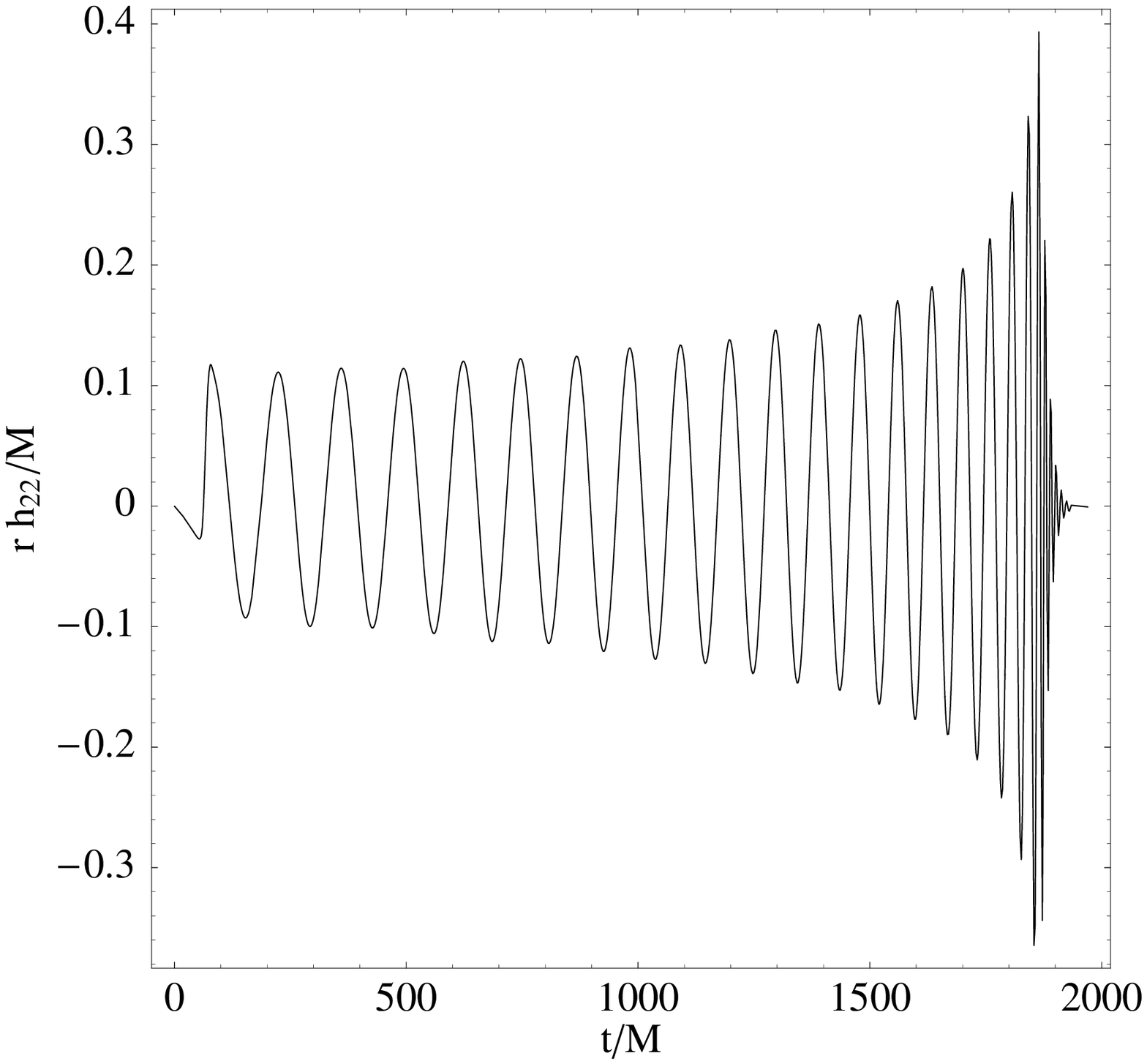}
\end{center}
\caption{Left panel: Coordinate tracks of the puncture location of one black hole
for simulations with grid sizes of the innermost boxes of $\{64^3,72^3,80^3\}$, starting
at a coordinate separation of $D=12 M$. 
Only in the last few orbits small differences are visible 
between the three runs discernable. Right panel: the waveform $h_+$, rescaled with the extraction radius.
Figure taken from \cite{Husa2007a}.
}
\label{fig:tracks_strain}
\end{figure}

In order to overcome phase inaccuracies in long evolutions, spectral
methods have been suggested and significant progress has been made
by the Caltech-Cornell group \cite{Scheel-etal-2006:dual-frame}.
The Jena group has recently obtained good results with 6th order
accurate finite differencing methods \cite{Husa2007a}, which has enabled us to perform highly accurate
evolutions over nine orbits (see fig. (\ref{fig:tracks_strain}) -- units in the figures
are given in units of the total initial black hole mass $M$) 
and perform a detailed comparison with predictions from PN 
approximations \cite{Husa2007a,Husa:2007ec,Hannam:2007ik}. 
For an extensive overview on PN approximations see \cite{Blanchet02}.

\begin{figure}[h!]
\begin{center}
\includegraphics[width=8cm,height=5cm]{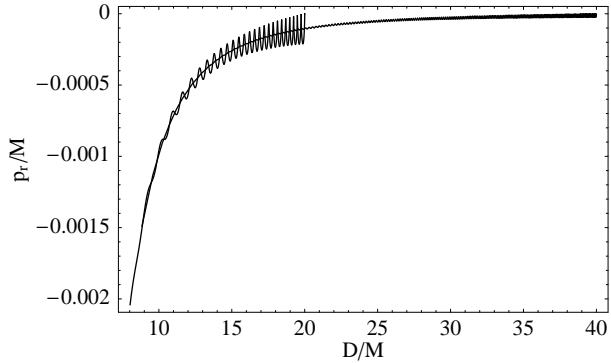}
\end{center}
\label{fig:eccentricity_decay}\caption{The radial momentum component is
  plotted versus separation
  for PN-inspirals starting from $D=20M$ and $D=40M$.
  A separation of $D=20M$ is clearly not sufficient to produce non-eccentric
  inspiral parameters,
  since  small oscillations can still be seen at $D=11M$, while for $D=40M$
  the initial eccentricity has essentially decayed away.
Figure taken from \cite{Husa:2007ec}.
}
\end{figure}

\begin{figure}[t!]
\begin{center}
\includegraphics[height=3.8cm]{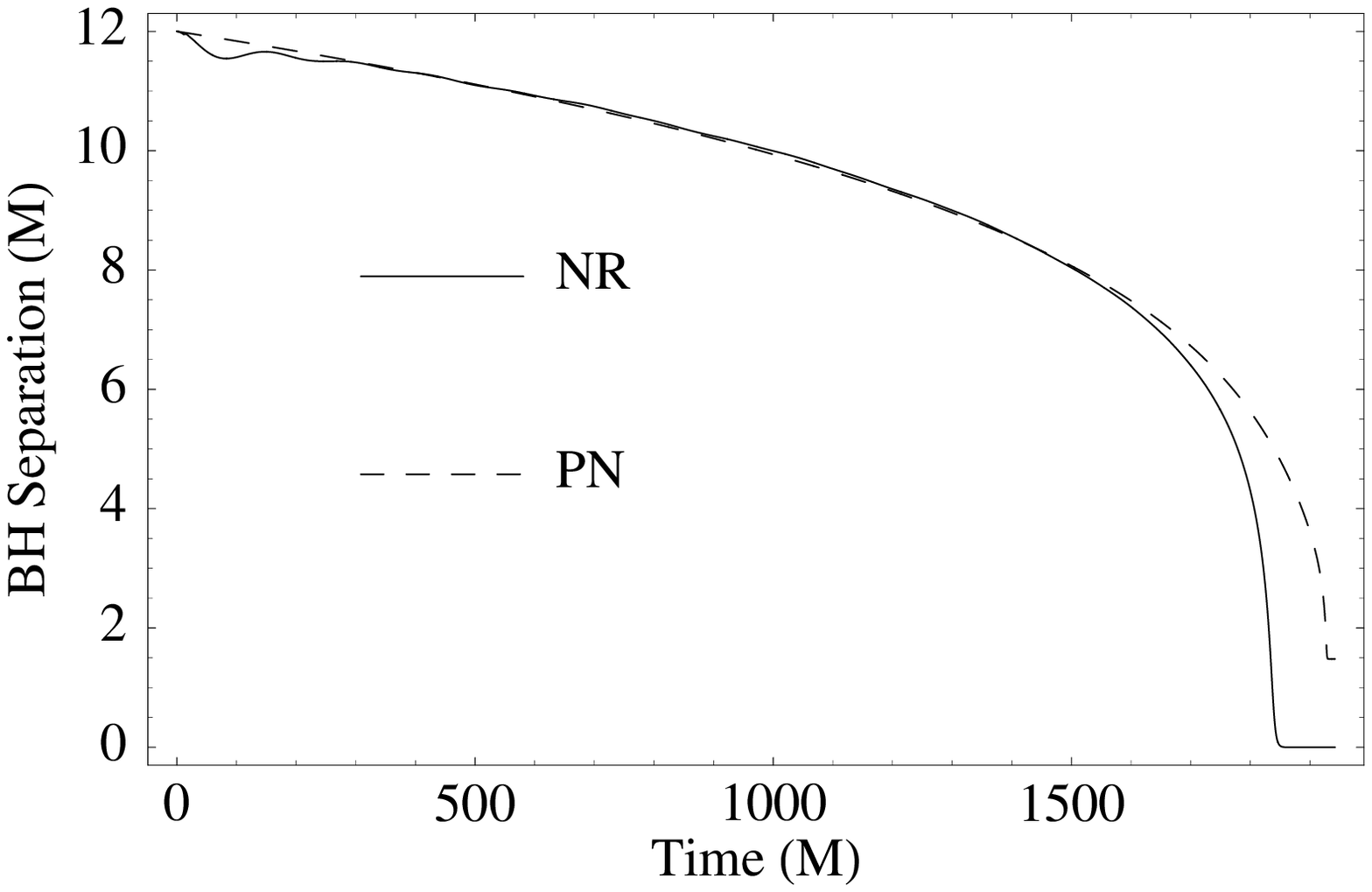}
\includegraphics[height=3.8cm]{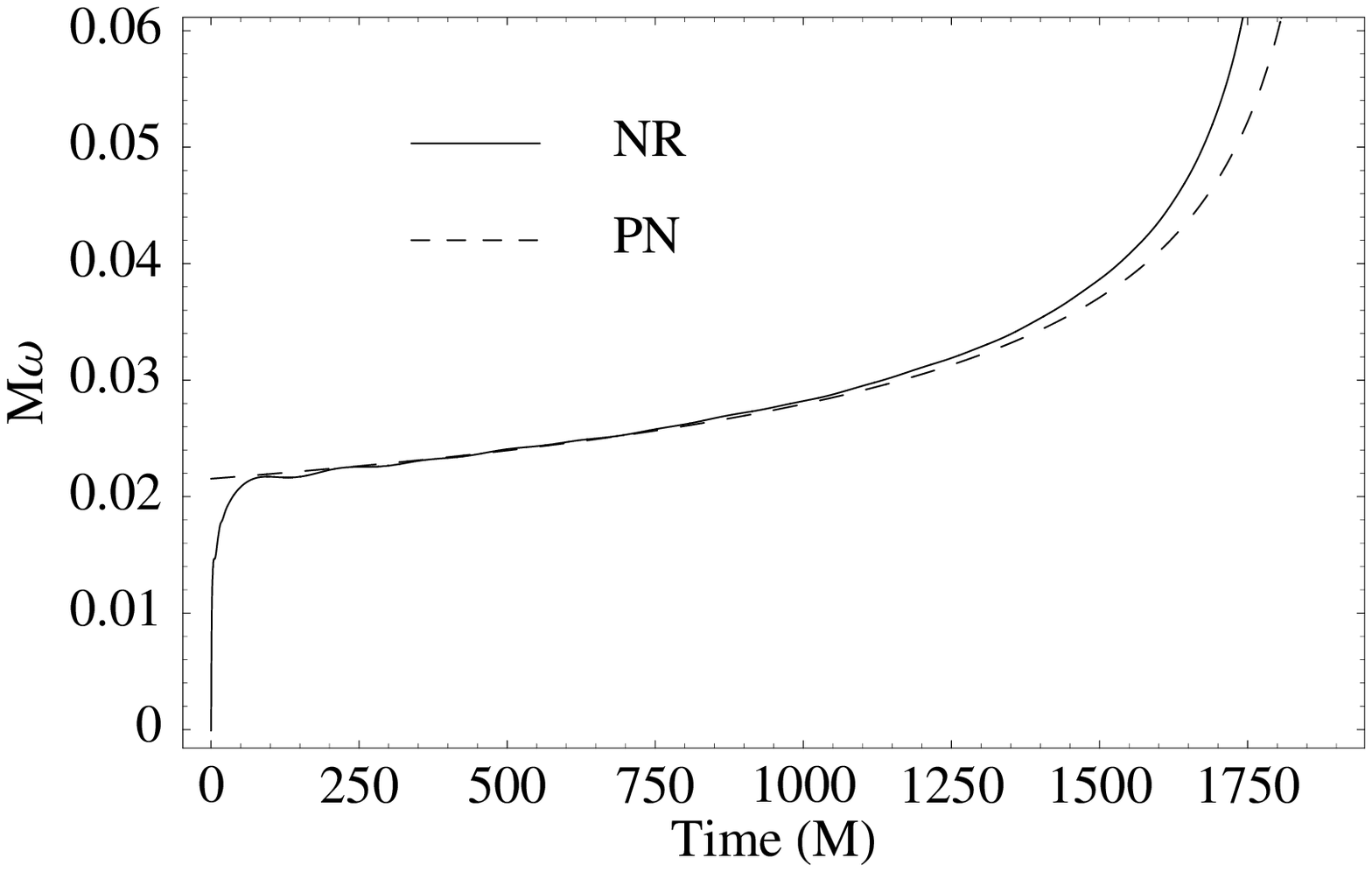}
\end{center}
\caption{Orbital coordinate motion of a 9-orbits numerical relativity evolution compared with a 
PN evolution with the same initial parameters. In both panels the PN evolution is drawn as a dashed line. 
Top panel: the separation of the black holes
(the puncture position in the full NR case).
Bottom panel: the coordinate angular velocity.
Figure taken from \cite{Hannam:2007ik}.}
\label{fig:tracks}
\end{figure}

A long standing problem has been the specification of non-eccentric initial 
data for numerical relativity simulations, which model inspiraling compact objects
which have shed the eccentricity of their orbit through gravitational radiation.
In  \cite{Husa:2007ec} we show, that at least for the nonspinning equal mass case the initial
momenta can be taken from a PN prescription. Part of the reason why this works
is that the coordinate gauge used for numerical relativity agrees with the so-called ADMTT gauge
adopted for PN calculations to excellent accuracy until very late in the inspiral, see fig.
(\ref{fig:tracks}). The PN calculation is performed such that all initial eccentricity is shed during
the first few hundred orbits, as shown in fig. (\ref{fig:eccentricity_decay}).

\begin{figure}[h!]
\begin{center}
\resizebox{0.5\columnwidth}{!}{%
  \includegraphics{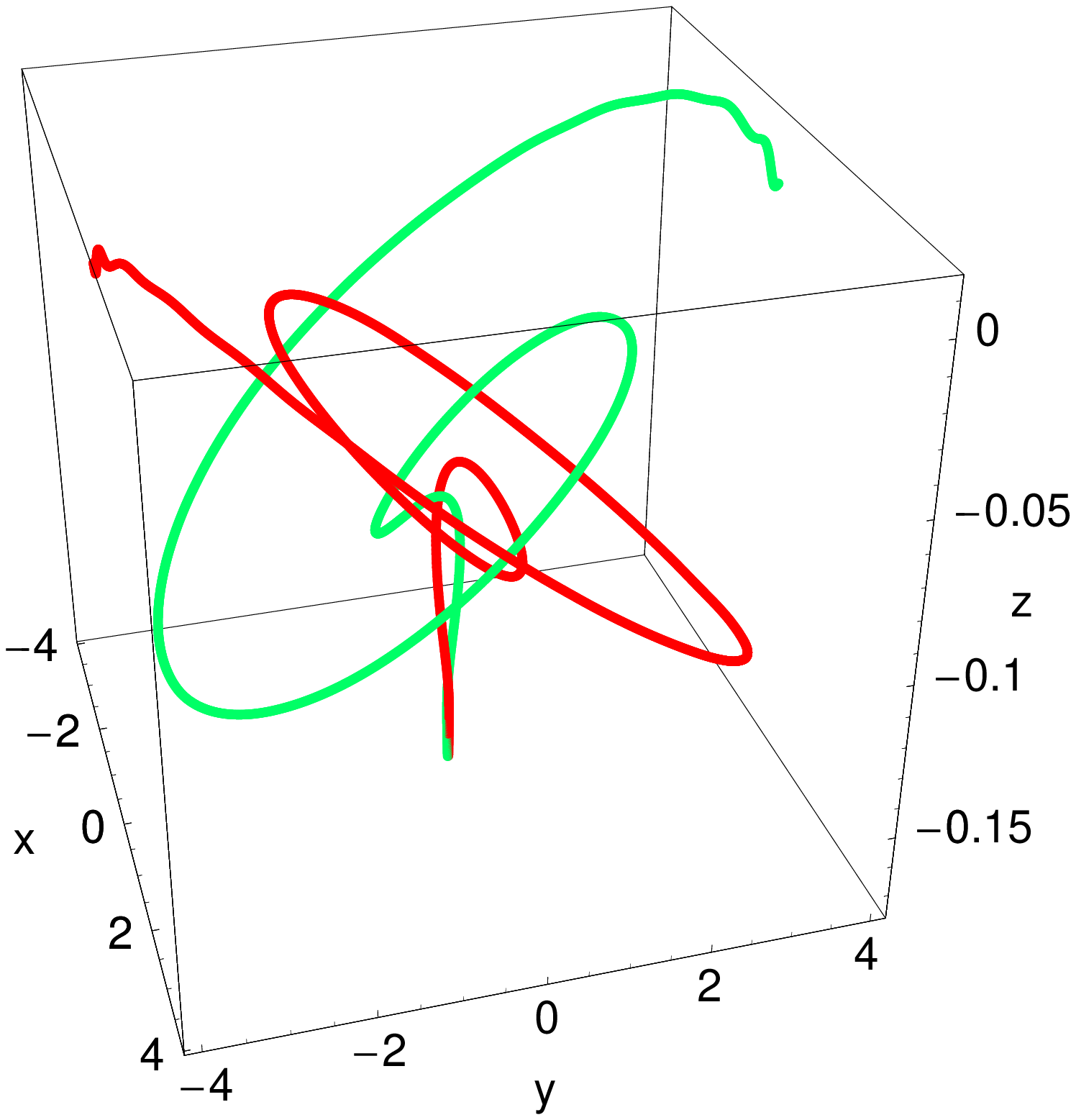} }
\end{center}
\caption{ Coordinate positions of the black-hole punctures for
model MII from \cite{Gonzalez:2007hi} up to $t = 180$. 
The black holes move out of the original plane
and after merger the final black hole receives a kick in the negative $z$-direction.
Figure taken from \cite{Gonzalez:2007hi}.}
\label{fig:fig3}
\end{figure}

A particular focus of the last few months
has been the so-called recoil or rocket
effect due to ``beamed'' emission of gravitational radiation
\cite{Bonnor1961,Peres1962,Bekenstein1973}.
By momentum conservation, radiation of energy in a preferred direction
corresponds to a loss of linear momentum and the black hole that results from
the merger thus recoils from the center-of-mass frame with speeds of up to several
thousand km/s. The velocity of this ``kick'' depends on the configuration of the
system ({\it e.g.} the mass ratios and spins) and details of the merger dynamics,
but not on the total mass (velocity is dimensionless in geometric units).

From an astrophysical point of view, the recoil effect is particularly interesting
for massive black holes with masses $> 10^5 \,M_{\odot}$, which exist at the
center of many galaxies and may have a substantial impact on the structure and
formation of their host galaxies.
The largest recoil effects have so far been found \cite{Gonzalez:2007hi,Campanelli:2007cg}
for a particularly simple configuration: equal mass black holes with (initially)
anti-aligned spins in the orbital plane.
Such large kicks are on the order of 1\% of the speed of light, and
larger than the escape velocity of about
$2000\,$km/s of giant elliptical galaxies. Smaller but still significant kick
velocities have been found for several different types of black hole
configurations \cite{Baker2006b,Herrmann2006,Gonzalez06tr,Herrmann:2007ac,Koppitz2007,Campanelli:2007ew}.

Many challenges remain in the binary black hole problem: extreme mass ratios, combined, say, with large spins
may remain problematic for some time; incorporating realistic matter models adds a wealth of new problems.
In order to perform large parameter studies, significant further optimizations
for current codes are probably required, along with further mathematical insights and a better understanding of 
the general relativity aspects
of the methodological foundations of the field.
I believe that with the ``binary black hole revolution'' the field of numerical relativity
has found a new beginning rather than come to its end -- 
and the study of colliding compact objects in particular will remain fruitful scientific 
ground for some time.

\begin{acknowledgement}
I am grateful to Mark Hannam, Norbert Lages and Christof Gattringer for reading the manuscript and
identifying some misprints and unclear points.
\end{acknowledgement}
\bibliographystyle{epj} 
\bibliography{refs}
\end{document}